%% file: main.tex
\documentclass[twocolumn]{aastex63}
\usepackage{amsmath}
\usepackage{xcolor}

\shorttitle{Search for PeV Gamma-Ray Emission with IceCube}
\shortauthors{M. G. Aartsen et al.}
\let\tablenum\relax
\usepackage{siunitx}
\begin{document}

\title{Search for PeV Gamma-Ray Emission from the Southern Hemisphere with 5 Years of Data from the IceCube Observatory}

\input{authors.tex}
\begin{abstract}

The measurement of diffuse PeV gamma-ray emission from the Galactic plane would provide information about the energy spectrum and propagation of Galactic cosmic rays, and the detection of a point-like source of PeV gamma rays would be strong evidence for a Galactic source capable of accelerating cosmic rays up to at least a few PeV. This paper presents several un-binned maximum likelihood searches for PeV gamma rays in the Southern Hemisphere using 5 years of data from the IceTop air shower surface detector and the in-ice array of the IceCube Observatory. The combination of both detectors takes advantage of the low muon content and deep shower maximum of gamma-ray air showers, and provides excellent sensitivity to gamma rays between $\sim$0.6~PeV and 100~PeV. Our measurements of point-like and diffuse Galactic emission of PeV gamma rays are consistent with background, so we constrain the angle-integrated diffuse gamma-ray flux from the Galactic Plane at 2~PeV to \num{2.61e-19}~\si{cm^{-2}s^{-1}TeV^{-1}} at 90\% confidence, assuming an E$^{-3}$ spectrum, and we estimate 90\% upper limits on point-like emission at 2 PeV between 10$^{-21}$ - 10$^{-20}$~\si{cm^{-2}s^{-1}TeV^{-1}} for an E$^{-2}$ spectrum, depending on declination. Furthermore, we exclude unbroken power-law emission up to 2~PeV for several TeV gamma-ray sources observed by H.E.S.S., and calculate upper limits on the energy cutoffs of these sources at 90\% confidence. We also find no PeV gamma rays correlated with neutrinos from IceCube’s high-energy starting event sample. These are currently the strongest constraints on PeV gamma-ray emission.

\end{abstract}


\section{Introduction}
\label{sec:intro}

Cosmic rays arriving at Earth approximately follow a power-law energy spectrum over eleven orders of magnitude, from 1 GeV to 100~EeV, with a slightly changing spectral index and only a few notable features. The softening of the spectrum at the `knee' at around 3~PeV and hardening of the spectrum at the `ankle' at around 3~EeV are the most prominent features of the spectrum. It is generally believed that the Galactic contribution to the cosmic-ray flux begins decreasing at the knee but extends up to the ankle, where the extra-Galactic population is responsible for the spectral hardening~\citep{Gaisser:2006}. However, this belief remains unsubstantiated since Galactic sources capable of accelerating cosmic rays above a PeV have not been identified yet. Cosmic-ray interactions with the gas near the accelerator produce neutrinos and gamma-ray photons. Unlike cosmic rays, neutrinos and gamma rays are unaffected by magnetic fields and are thus critical towards the identification of these accelerators.
Further, the cosmic rays which escape the local environment of their sources propagate through the Galaxy and interact with interstellar gas.  The observable emission of neutrinos and gamma rays is expected to peak along the Galactic plane, where most of the interstellar gas is concentrated~\citep{Kalberla:2009}. The measurement of this diffuse emission can provide information about the cosmic-ray diffusion processes and gauge the cosmic-ray spectrum at Galactic locations other than the Earth (\citealt{Gaggero:2015a} and \citealt{Acero:2016}).

The IceCube Observatory has observed an isotropic flux of astrophysical neutrinos (\citealt{Aartsen:2013d}, ~\citealt{Aartsen:2015c}, ~\citealt{Aartsen:2015d}) but no point-like sources have been resolved so far, except for recent strong indications of an extra-Galactic source based on multi-messenger observations~\citep{Aartsen:2018}.
A recent study using neutrino data from both IceCube and ANTARES constrained the Galactic plane contribution to the isotropic flux to no more than 8.5\%~\citep{Albert:2018icant}.

A complementary PeV gamma-ray search in the energy range of $\sim$0.6~PeV to 100~PeV is made possible by the presence of the surface air shower component of the IceCube Observatory. 
 In this energy regime, gamma rays can only be observed over Galactic distances due to the high cross-section for pair production with the cosmic microwave background (CMB) radiation field~\citep{Protheroe:1996}. Therefore, the measurement of PeV gamma rays can further constrain the Galactic contribution to the observed astrophysical neutrino flux. As the sole experiment to-date sensitive to PeV gamma rays in the Southern Hemisphere, the IceCube Observatory offers a unique window to high-energy processes in our Galaxy.

This paper will summarize the PeV gamma-ray measurements of the IceCube Observatory, and is a follow up of the previous study by \citet{Aartsen:2013}, who used data taken over one year with a partial configuration of IceCube consisting of 40 strings (IC-40). Here, we analyze five years of data from the completed observatory with 86~strings and include inclined events recorded only by the surface array which significantly increases the detector acceptance over the entire field of view of $-90^{\circ} \leq \delta \leq -53^{\circ}$ (declination).

In the first part of this analysis, we obtain an air shower event sample rich in gamma-ray candidates by exploiting the key differences between air showers of cosmic-ray and gamma-ray origin. The most effective discriminator is the number of muons produced in the air shower.  Muons are created in gamma-ray air showers from muon pair production as well as the decay of photo-produced pions and kaons (\citealt{Drees:1989},~\citealt{Halzen:2009}).  However, these processes are much less frequent than muon production from nucleus-nucleus interactions in hadronic showers.  From CORSIKA~\citep{Heck:1998} simulations utilizing hadronic interaction models FLUKA~\citep{Battistoni:2007} and SYBILL~2.1~\citep{Ahn:2009}, we find that 1~PeV vertical proton showers contain roughly ten times the number of 1~GeV muons at the IceTop surface compared to 1~PeV vertical gamma-ray showers. This ratio increases to roughly a hundred for muons with energy greater than 460~GeV at the surface. The in-ice array of IceCube is sensitive to muons highly collimated around the shower axis with energies greater than $\sim$460~GeV at the surface. While the surface array is crucial to measure the energy deposited in the electromagnetic part of the shower, it is also sensitive to lower energy muons arriving far from the shower core. Additionally, the shower maximum from gamma-ray primaries occurs on average deeper in the atmosphere, resulting in a younger shower age~\citep{Risse:2007}. The stage of longitudinal shower development can be assessed from the electromagnetic component observed by the surface array. The gamma-hadron discrimination method is detailed in Section~\ref{sec:classifier}.

In the second part of this analysis, we search for point-like sources of PeV gamma rays in the Southern Hemisphere using the event sample containing gamma-ray-like events. The current generation of ground-based air Cherenkov detectors have uncovered a wealth of Galactic TeV gamma-ray sources (e.g.~\citealt{Abeysekara:2017},~\citealt{Benbow:2017},~\citealt{Carrigan:2013}). Of particular interest in this analysis are the results of the High Energy Spectroscopic System (H.E.S.S.), which has a field of view that overlaps with that of IceTop. H.E.S.S. is the only experiment to detect sources that show no evidence of a cutoff at TeV energies in a location testable by this analysis.  These sources include Pulsar Wind Nebulae (PWN), Supernova Remnants (SNR), and several other unclassified sources~\citep{Carrigan:2013}.  We search for emission spatially correlated with these known TeV gamma-ray sources in addition to an unbiased search for PeV gamma-ray sources across the entire analysis field of view.  We also search for PeV counterparts to the IceCube neutrino events with a high likelihood of astrophysical origin~\citep{Aartsen:2015b}, a component of which may be of Galactic origin (\citealt{Joshi:2013aua}, \citealt{Ahlers:2013xia}, \citealt{Kachelriess:2014oma}, \citealt{Ahlers:2015moa}). The search methods are described in Section \ref{sec:psllhmethod} and the results of the point source searches are presented in Sections \ref{sec:psresult}, \ref{sec:hessresult}, and \ref{sec:heseresult}.

Diffuse gamma-ray emission from the Galactic plane has been measured by ground-based air/water Cherenkov observatories up to $\sim$10 TeV (\citealt{Hunger:1997we}, \citealt{Aharonian:2006}, \citealt{Abdo:2008}, \citealt{Ackermann:2012pya}). In the Northern hemisphere, CASA-MIA~\citep{Borione:1998} has placed upper limits on a diffuse flux from the Galactic plane between 140~TeV and 1.3~PeV, while KASCADE-Grande has reported limits on an isotropic diffuse flux of gamma rays from 100~TeV to 1~EeV~\citep{Apel:2017ocm}.  The IC-40 analysis~\citep{Aartsen:2013} produced the sole limit on the PeV flux from a section of the Galactic plane visible in the Southern Hemisphere.  In the third part of this analysis, we search for a diffuse flux from the Galactic plane within the field of view of IceTop ($\delta \leq -53^{\circ}$). We use the neutral pion decay component of the Fermi-LAT diffuse emission model~\citep{Ackermann:2012} as a spatial template in a maximum likelihood analysis described in Section \ref{sec:diffusellhmethod}). The result of this search is presented in Section \ref{sec:gpresult}.

\section{The IceCube Observatory}
\label{sec:detector}

\begin{figure}
\plotone{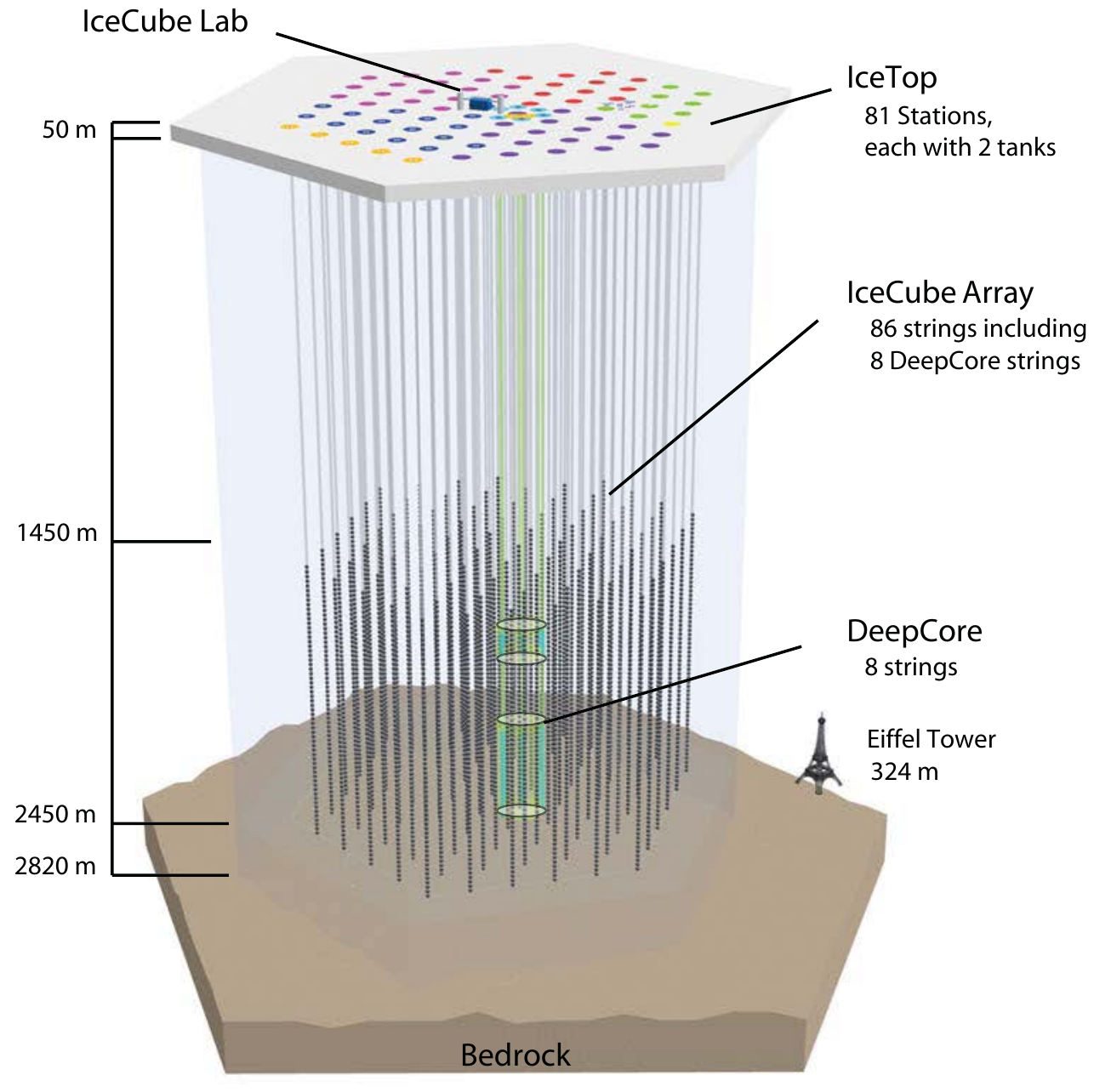}
\caption{A schematic of the entire IceCube observatory~\citep{Aartsen:2017d}.  The surface array (IceTop) and the in-ice array are shown, along with the in-ice subarray DeepCore.}
\label{fig:icecube}
\end{figure}

Located at the geographic South Pole, the IceCube observatory (sketched in Figure~\ref{fig:icecube}) consists of two major components - an in-ice array and a companion surface array known as IceTop. The in-ice array is capable of detecting neutrinos in the energy range of 100~GeV to EeV and high energy muons originating in the cosmic-ray showers, whereas the surface array is designed to detect air showers from cosmic rays in the energy range of 300~TeV to EeV. The IceCube observatory~\citep{Aartsen:2017d} was completed in 2010 following seven years of construction.

The cubic kilometer in-ice array is comprised of a total of 5,160 optical sensors, or digital optical modules (DOMs), organized on 86 cables installed in the ice between depths of 1450 m and 2450 m.  Each DOM contains a 10-inch Hamamatsu photomultiplier tube in addition to electronic boards necessary for triggering, digitization, and readout~\citep{Abbasi:2009}. The in-ice array detects the Cherenkov photons emitted by relativistic charged particles traversing the array.

\begin{figure}
\plotone{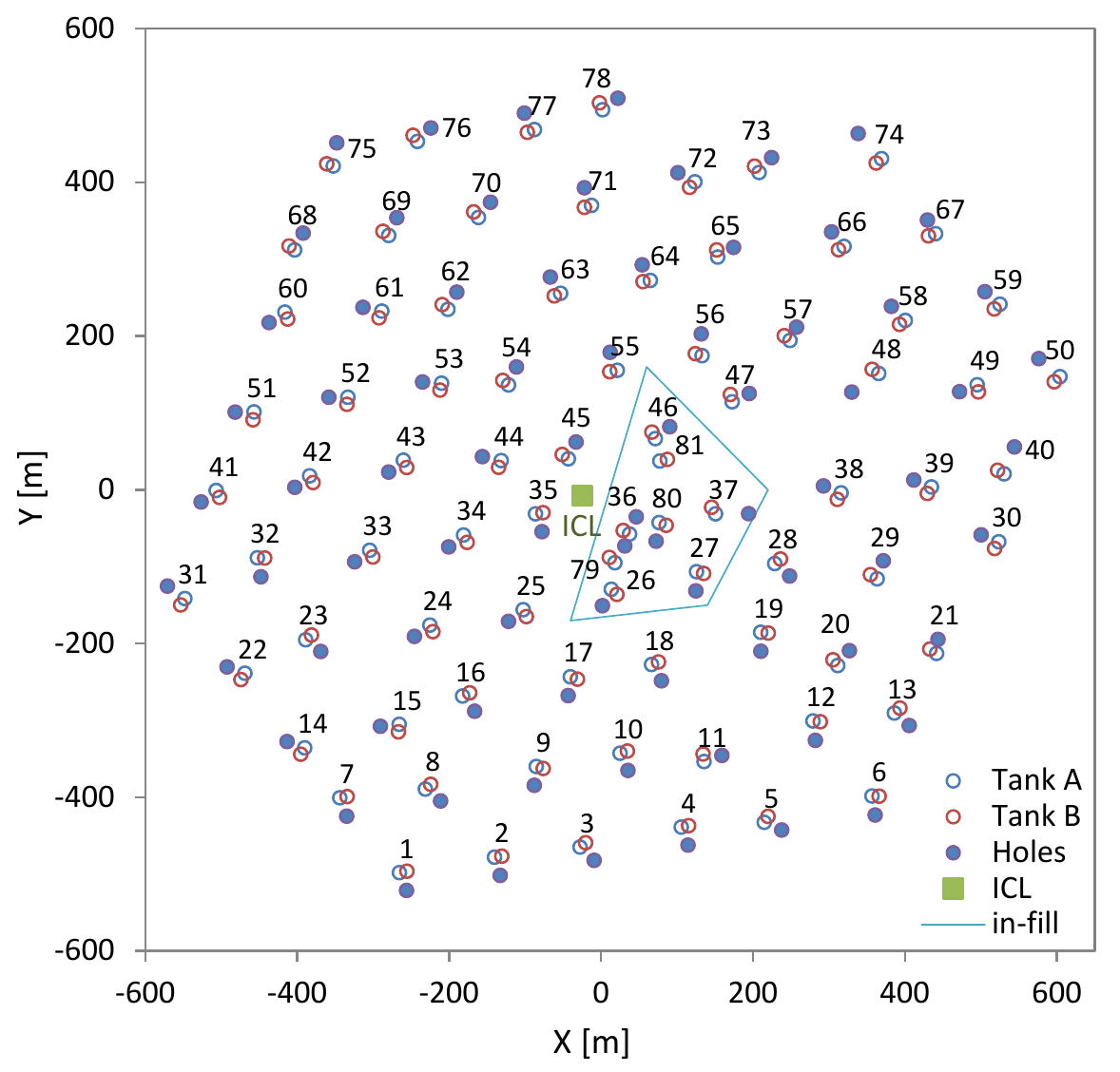}
\caption{IceTop array geometry with locations for 81 stations and in-ice string holes. The stations that form the denser in-fill array for lower energy showers are also demarcated.}
\label{fig:icetop}
\end{figure}

The surface array of the IceCube observatory, IceTop~\citep{Abbasi:2013}, is located on top of the ice sheet at an altitude of 2835 meters above sea level. The geometry of IceTop is displayed in Figure~\ref{fig:icetop}.  IceTop is comprised of 81 stations, where a station is defined as two tanks filled with ice that are situated above a subset of IceCube strings.  Each tank contains two DOMs embedded in ice and configured with different PMT gains in order to increase the dynamic range of signal measurement. In this way, each tank is capable of detecting the Cherenkov radiation from muons and other charged particles traversing it.  IceTop triggers on extensive air showers by measuring the Cherenkov light inside the tanks emitted by air shower particles or by secondary particles from their interactions in ice. At the surface, a typical cosmic-ray muon has an energy of a few GeV. Such minimum ionizing muons, passing through a tank, will deposit roughly the same amount of energy depending on their path length inside the tank. Every DOM's charge spectrum from single muons, obtained from low-energy cosmic ray showers, is fitted to find the charge value corresponding to vertical muons. Thus all IceTop DOMs are calibrated to convert their charge value to a standard ’Vertical Equivalent Muon’ (VEM) unit.

\input{event_numbers_table.tex}

In this analysis, both the surface and in-ice arrays are used to discriminate gamma-ray induced air showers from cosmic-ray induced air showers. Reconstructions using IceTop signals for an air shower provide the energy proxy, arrival direction, and a primary mass sensitive parameter (see Section \ref{sec:llhratio}), while the in-ice array provides an estimate of the energy deposited by $\sim$TeV muons for air showers whose axis passes through the deep ice detector.

\section{Dataset Construction}
\label{sec:qc}

Five years of experimental data is used in this analysis, collected between May 2011 and May 2016. We use a machine learning classifier to separate gamma-ray signal from cosmic-ray background. During the training of this classifier, 10\% of the observed data is used to model background, leaving 90\% available for the final analysis. Since the expected fraction of gamma rays in the air shower data is very low ($\mathcal{O}(10^{-4})$), we use data in lieu of Monte-Carlo simulations as a proxy for the cosmic-ray background. This greatly reduces the systematic dependencies inherent to simulation such as the choice of hadronic interaction model, atmospheric model, cosmic-ray composition model, and snow height averaging. The total livetime of the data used for each year in the final analysis is listed in Table~\ref{table:event_numbers_table}. To model signal, Monte-Carlo simulations of gamma-ray air showers were produced using CORSIKA version 7.37, with high-energy hadronic interactions treated with SYBILL~2.1 and low-energy hadronic interactions modeled using FLUKA.  80\% of the gamma-ray Monte-Carlo was used in the training of the event classifiers, with the remaining 20\% withheld to test the final analysis performance. The simulated gamma-ray showers were weighted to a power-law spectrum, with the choice of spectral index depending upon the source hypothesis. Simulations and data were treated identically with regards to event processing, event selection, and event reconstruction.

\subsection{Air Shower Event Reconstruction}
An event is recorded in IceTop whenever at least three stations (six tanks) are hit by a shower front within a time interval of 6 $\mu$s. Then the data recorded from the collection of tanks for each event is filtered to remove uncorrelated background particle hits. The simultaneous reconstruction of shower size and arrival direction is carried out through the maximization of likelihood functions describing the lateral distribution of the signal and the shower front shape~\citep{Abbasi:2013}. The signal charge distribution, $S$, around the shower core in the shower frame of reference is known as the lateral distribution function (LDF). The LDF chosen to describe air shower events in IceTop is an empirically derived double logarithmic parabola. At a lateral distance $R$ from the shower axis, the charge expectation $S$ in an IceTop tank is defined as%
\begin{equation}
\label{eq:shower_size}
S(R) = S_{125}\left(\frac{R}{125~\mathrm{m}}\right)^{-\beta-\kappa \log_{10}(R/125~\mathrm{m})}
\end{equation}%
where $S_{125}$, also referred to as shower size, is the fitted signal strength at a reference distance of 125 meters, $\beta$ is the fitted slope parameter correlated with shower age, and $\kappa$ = 0.303 is a constant determined through simulation studies. The shower size, $S_{125}$, is proportional to the energy of the primary particle, and is converted to a reconstructed energy, E$_{\mathrm{reco}}$, using parameterization obtained from cosmic-ray simulations~\citep{Rawlins:2015nvu}.

\begin{figure}[htb!]
\plotone{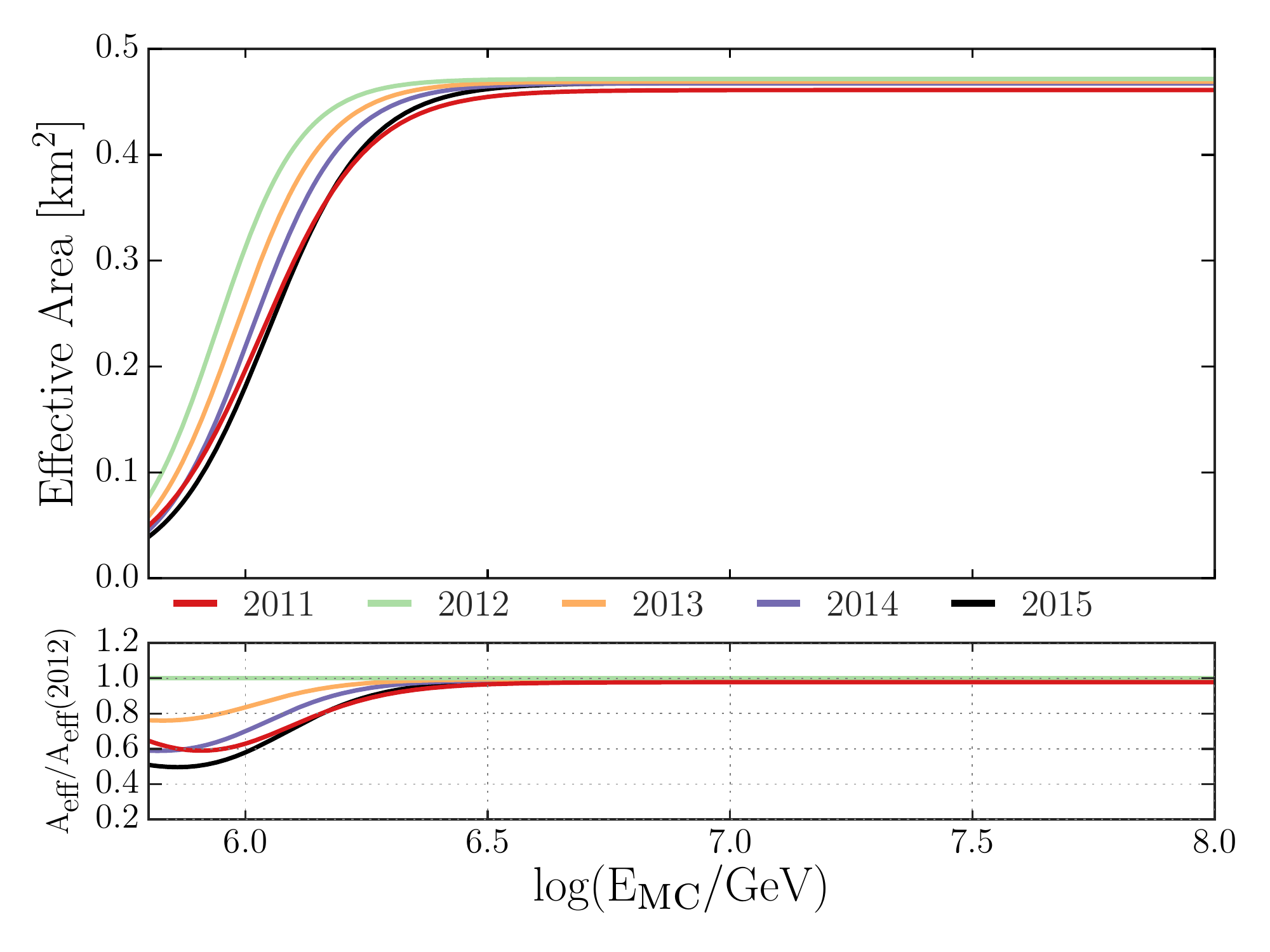}
\caption{The effective area of IceTop to gamma rays simulated with snow heights from each year of the data-taking period of the analysis. All cuts listed in Section \ref{sec:qc} were applied.}
\label{fig:eff_area_comp}
\end{figure}

Accumulation of snow on top of the IceTop tanks suppresses the electromagnetic portion of the air shower, requiring a correction factor to the signal expectation defined as%
\begin{equation}
\label{eq:snow_correction}
S^{\mathrm{corr}}_i = S_i~\mathrm{exp}\left(\frac{h^i_s}{\lambda_s\cos\theta}\right)
\end{equation}%
where $S^{\mathrm{corr}}_i$ is the corrected signal expectation, $h^i_s$ is the snow height above tank $i$, $\theta$ is the reconstructed zenith angle of the shower, and $\lambda_s$ is the effective absorption length in snow.  The value for $\lambda_s$ used for each analysis year is also listed in Table~\ref{table:event_numbers_table}. The effective absorption length was optimized by comparing each year's snow-corrected $S_{125}$ spectrum with the spectrum from the first year of operation with least amount of snow. Snow on top of IceTop tanks is constantly increasing at a rate of $\sim$20 cm per year, which significantly degrades the detector acceptance. In order to accurately account for this effect, the detector response was simulated for each year of data included in the analysis using the same set of CORSIKA gamma-ray showers.  For each dataset, the snow level on top of the tanks was set to the snow heights measured during the austral summer season.  Figure~\ref{fig:eff_area_comp} shows the effective area of IceTop to gamma rays for each year as a function of the primary energy, illustrating the event rate loss caused by increasing amounts of snow. The effective area is lower for the 2011 data year due to low statistics for events with less than eight stations because only one in three such events were being transmitted north from the South Pole in 2011.

\subsection{Quality Event Selection}
\begin{figure}[htb!]
\centering
\includegraphics[scale=0.35]{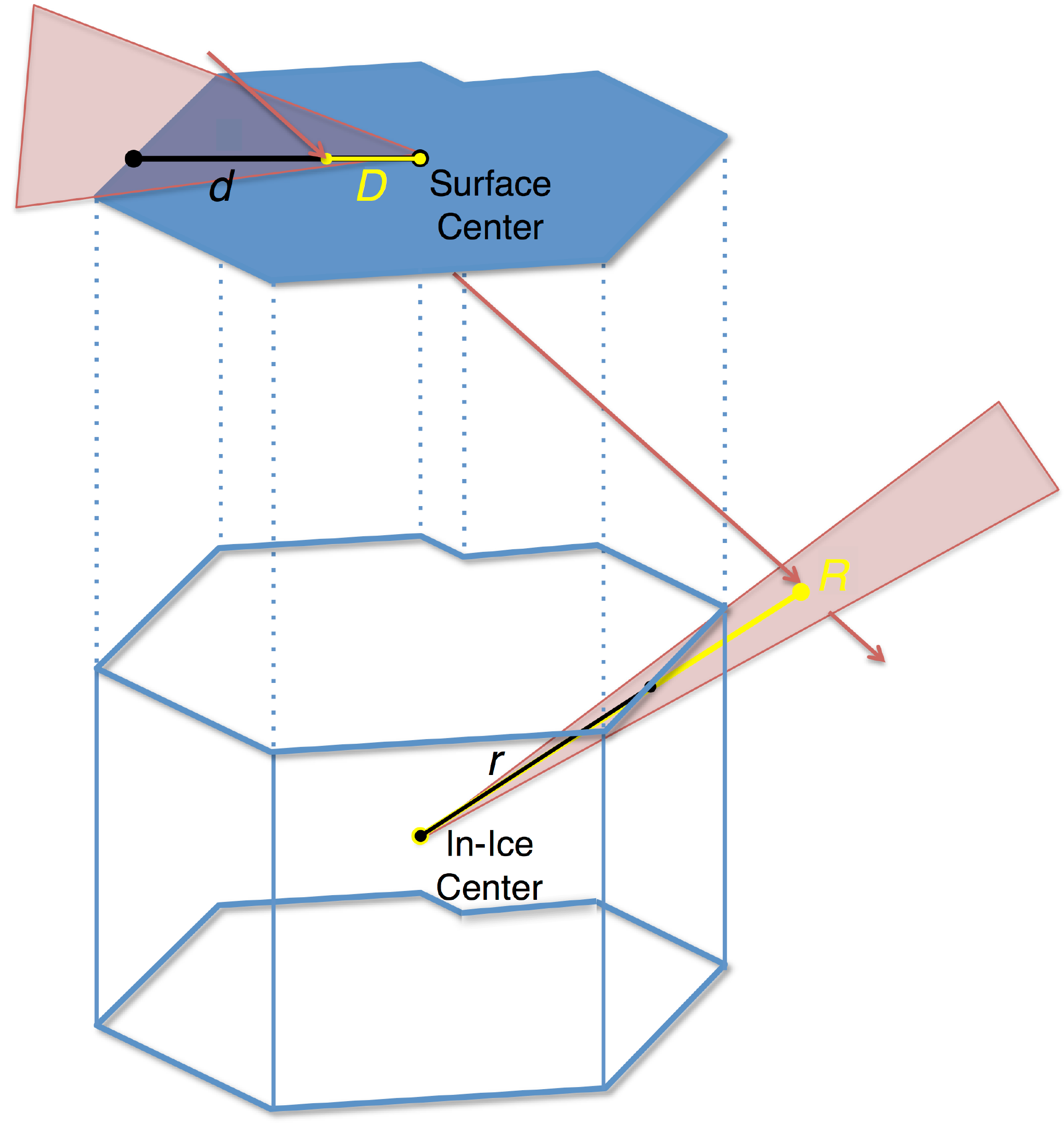}
\caption{Schematic diagram representing the parameters used in the calculation of containment in IceTop and the in-ice array.  For IceTop, $D$ and $d$ are the distances from the geometric center of the surface array to the shower core and the edge of the array in the direction of the shower core, respectively.  For the in-ice array, $R$ is the closest distance between the geometric center of the in-ice array and the reconstructed shower vertex while $r$ is the distance to the edge of the in-ice array along the same line.}
\label{fig:containment}
\end{figure}

To ensure good energy and direction reconstruction, the following quality cuts were applied:%
\begin{enumerate}
\item Events must trigger at least 5 IceTop stations (i.e., where both tanks in the station have DOMs with deposited charge within 1 $\mu$s).
\item The energy and directional reconstructions must converge.
\item The reconstructed core position must be contained within the surface array geometry. Specifically, the event must satisfy $D$/$d$ $<$ 1, where $D$ and $d$ are defined in Figure~\ref{fig:containment}.
\item The tank with the largest deposited charge must not lie on the edge of the IceTop array.
\item At least one IceTop tank must have a signal greater than 6 VEM.
\item The reconstructed zenith angles satisfy cos($\theta$)~$>$~0.8.
\item The reconstructed shower sizes satisfy $\log_{10}(S_{125})>-0.25$.
\end{enumerate}
We limit the event selection to E$_{reco}\leq$100 PeV, as extending to higher energies would have required additional higher energy gamma-ray simulations for a negligible improvement in sensitivity.

\begin{figure}[htb!]
\plotone{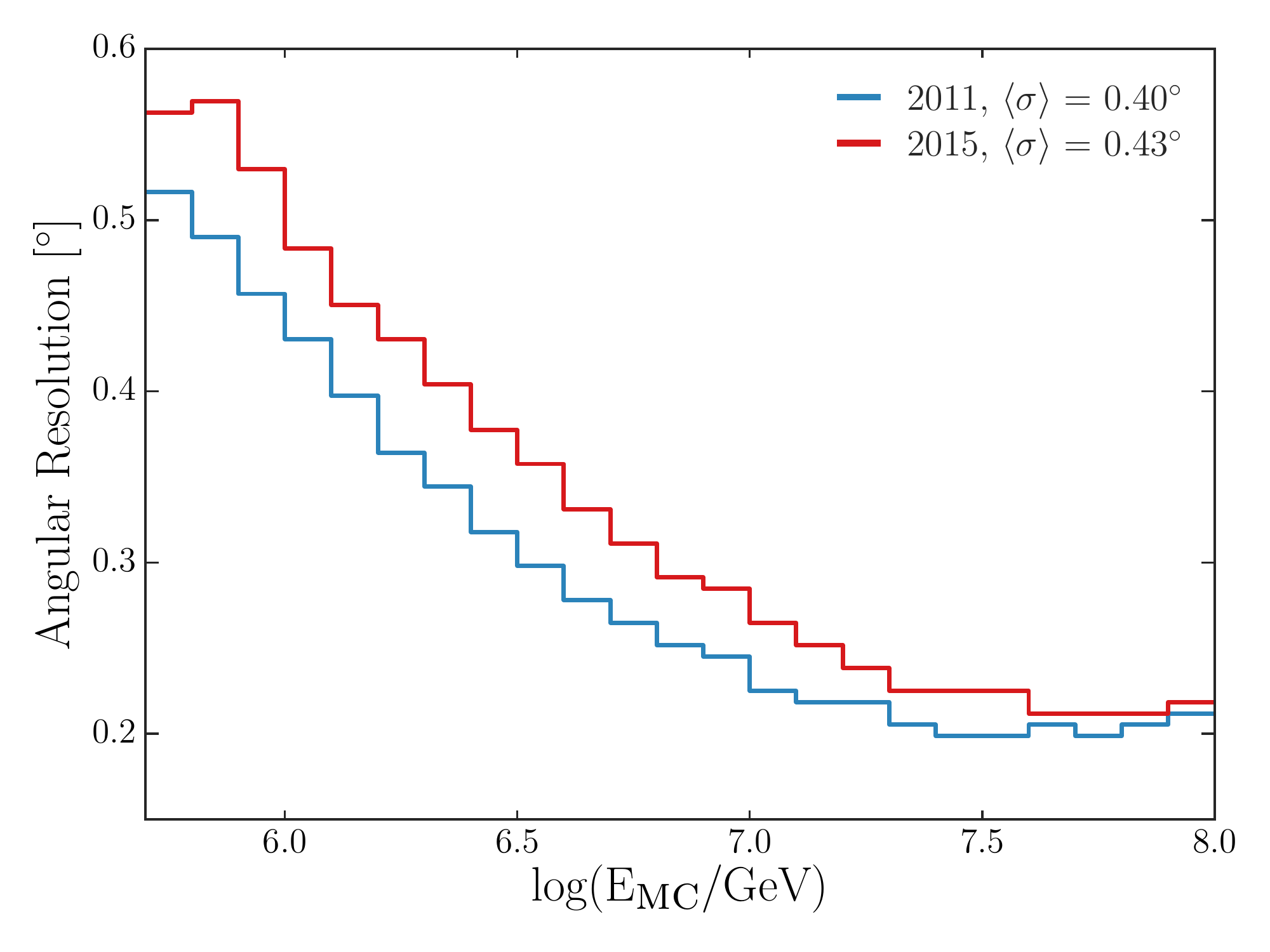}
\caption{68\% containment intervals of angular resolution to gamma rays from simulation using detector response with snow heights from the first and last year of the data-taking period of the analysis. $\langle\sigma\rangle$ denotes the median angular resolution assuming an E$^{-2.0}$ energy spectrum.}
\label{fig:ang_res}
\end{figure}

Angular resolution, defined as the angular radius that contains 68\% of reconstructed showers coming from a fixed direction, drives the sensitivity of searches for point-like and extended sources. Figure~\ref{fig:ang_res} shows the angular resolution of simulated gamma rays as a function of the true primary energy. The first and last years are shown to illustrate the impacts of the snow accumulation on the angular resolution, which amounts to a degradation of around 8\% on average.

\subsection{Discriminating Gamma-ray Showers}
\label{sec:classifier}
\begin{figure}[htb!]
\plotone{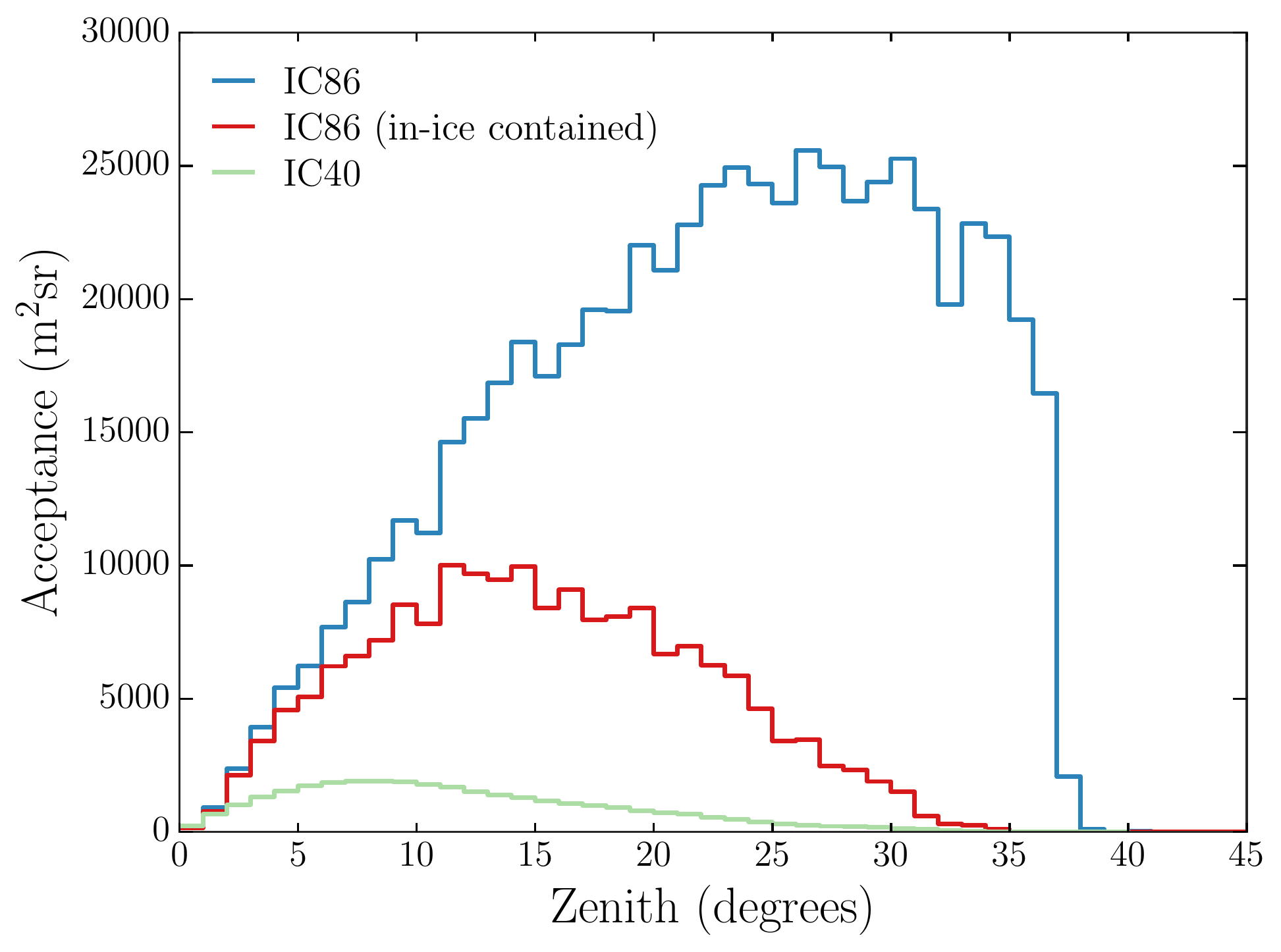}
\caption{The detector acceptance (effective area integrated over solid angle) as a function of the zenith angle for all gamma rays (blue line), the subset of gamma rays passing through the in-ice array (red line), and for the IC-40 analysis~\citep{Aartsen:2013}. The detector response for IC-86 gamma-ray simulations shown here were done using snow heights from October 2012. All three distributions were made assuming an E$^{-2.7}$ gamma-ray spectrum.}
\label{fig:acceptance}
\end{figure}

In order to extract all the shower information correlated with the type of the primary particle, both surface and in-ice components of the detector are utilized. Section \ref{sec:iniceclassifier} details the technique used to obtain clean data from the in-ice array which informs on the amount of high energy muons in the shower. Section \ref{sec:llhratio} describes the implementation of a new likelihood method which optimally retrieves information from the surface array correlated with the number of low-energy muons, shower age, and shower profile. IceTop based shower discrimination allows us to include showers which do not pass through the in-ice array in the analysis. While the loss of in-ice information for these showers reduces separation power, there is a large increase in detector acceptance at higher zenith angles, which is displayed in Figure \ref{fig:acceptance}. We combine the in-ice and surface components in a single classifier using machine learning as described in Section \ref{sec:rf}.

\subsubsection{High Energy Muons In Ice}
\label{sec:iniceclassifier}
Muons that have energy greater than $\sim$460~GeV at the surface are capable of generating light that can be detected by the photomultipliers of the in-ice array. This is the main parameter used to judge the hadronic content of air showers with trajectories that pass through the in-ice detector component. We use the total charge measured in-ice as a discriminating feature. In order to isolate the signal produced by muons, a cleaning procedure is applied to remove charge deposited by uncorrelated background particles. The cleaning procedure for events with or without an in-ice trigger are described below.

Nearest or next-to-nearest neighboring DOMs on the same string that have deposited charge (a `hit') within a time window of $\pm$1 $\mu$s are designated to be in Hard Local Coincidence (HLC). An in-ice trigger is defined as 8 or more such HLC hits in an event within a 5 $\mu$s time window. An in-ice trigger which falls between 3.5 $\mu$s and 11.5 $\mu$s after the start of an IceTop trigger is considered coincident, in which case hits outside of the coincident time window are removed.  Next, HLC hits are used as seeds in a hit selection algorithm which searches for single hits which are within 150 meters and 1 $\mu$s of each seed DOM. In an iterative manner, single hits which satisfy the criteria are included in the seeds and the search is performed again.  This is repeated three times.  The combined charge of the final selection of hits is used as a discriminating feature.

\begin{figure}[htb!]
\plotone{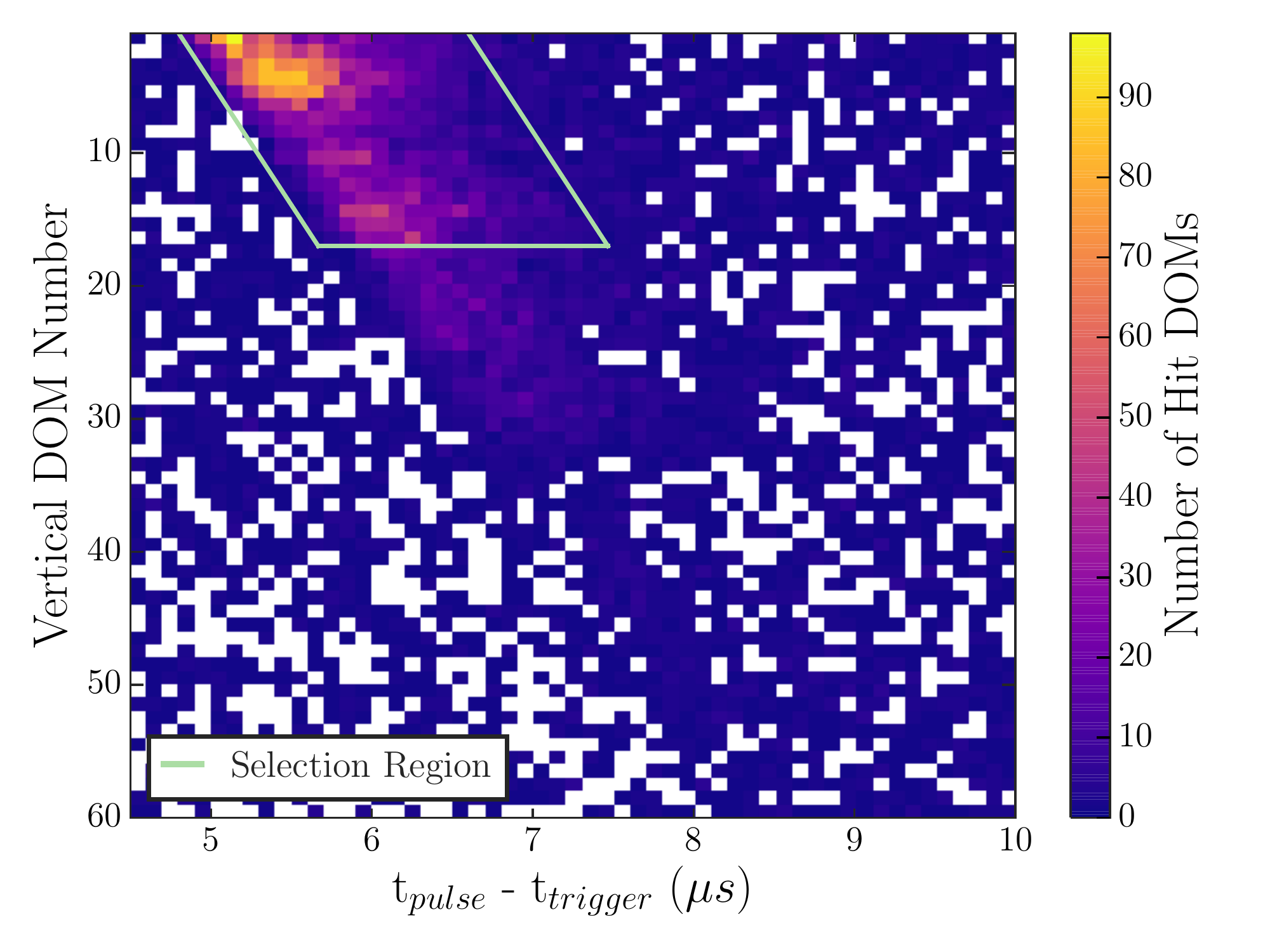}
\caption{The number of in-ice array hit DOMs binned in vertical DOM number and the hit time ($t_{\textrm{hit}}$) relative to the IceTop trigger ($t_{\textrm{trigger}}$), for a collection of experimental events that have no in-ice array HLC hits.  The vertical DOM number denotes the position of the hit DOM on its string; larger numbers are on deeper layers of the array.  The green box delimits the region within which charge is selected for event classification.}
\label{fig:slc_optimization}
\end{figure}

For those events which have only an IceTop trigger, simpler cleaning is applied.  For HLC hits, a time window selection is used to reduce the uncorrelated muon background, optimized to be between 3.5 $\mu$s and 9 $\mu$s after the trigger in IceTop. Hadronic showers may produce isolated hits in IceCube without an HLC flag.  This is illustrated in Figure~\ref{fig:slc_optimization}, which displays the number of single hits as a function of vertical DOM number (analogous to depth) and time with respect to the trigger for a collection of events which have no HLC hits.  There is a clear excess in hits near the top of the detector that is suitable for classification.  A selection window is optimized by a maximization of the ratio of signal and square root noise.  We define the front of the time window for charge selection as:%
\begin{equation}
\label{eq:time_window_front}
t_{\textrm{start}} = \frac{4.8 \mu\textrm{s} \ + d_{\textrm{DOM}}/c}{\cos(\theta)}
\end{equation}%
where $\theta$ is the reconstructed zenith angle of the event, $d_{\textrm{DOM}}$ is the depth of the hit in meters, and $c$ is the speed of light.  Hits are retained if they fulfill the following criteria, represented by the green box shown in Figure~\ref{fig:slc_optimization}:%
\begin{enumerate}
\item The hit is within 130 meters of the reconstructed shower axis.
\item The hit is within the top 16 layers of in-ice DOMs.
\item The hit has a time stamp $t$, such that $t_{\textrm{start}} < t < t_{\textrm{start}} + 1.8~\mu\textrm{s}$.
\end{enumerate}

\begin{figure*}[htb!]
\plottwo{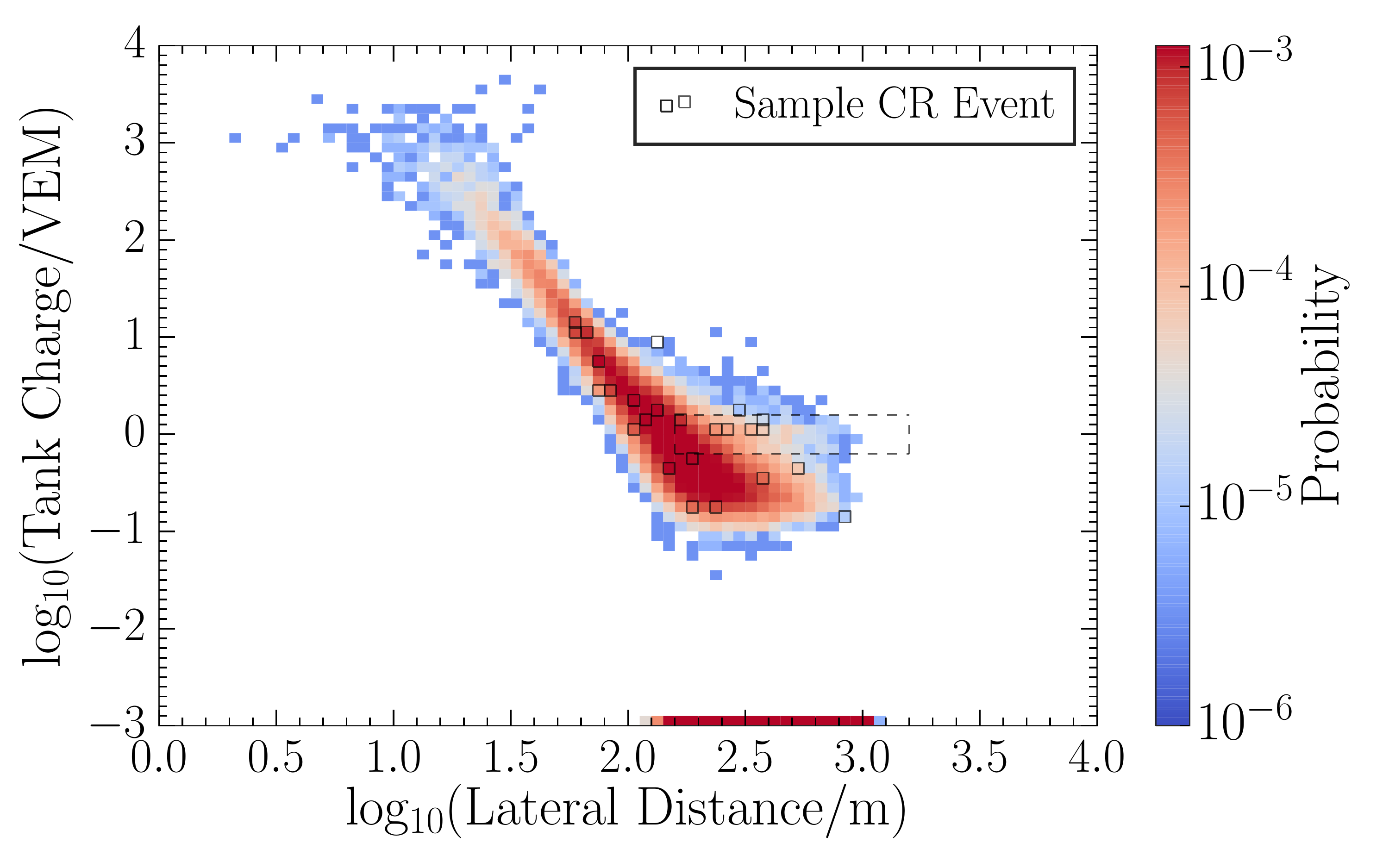}{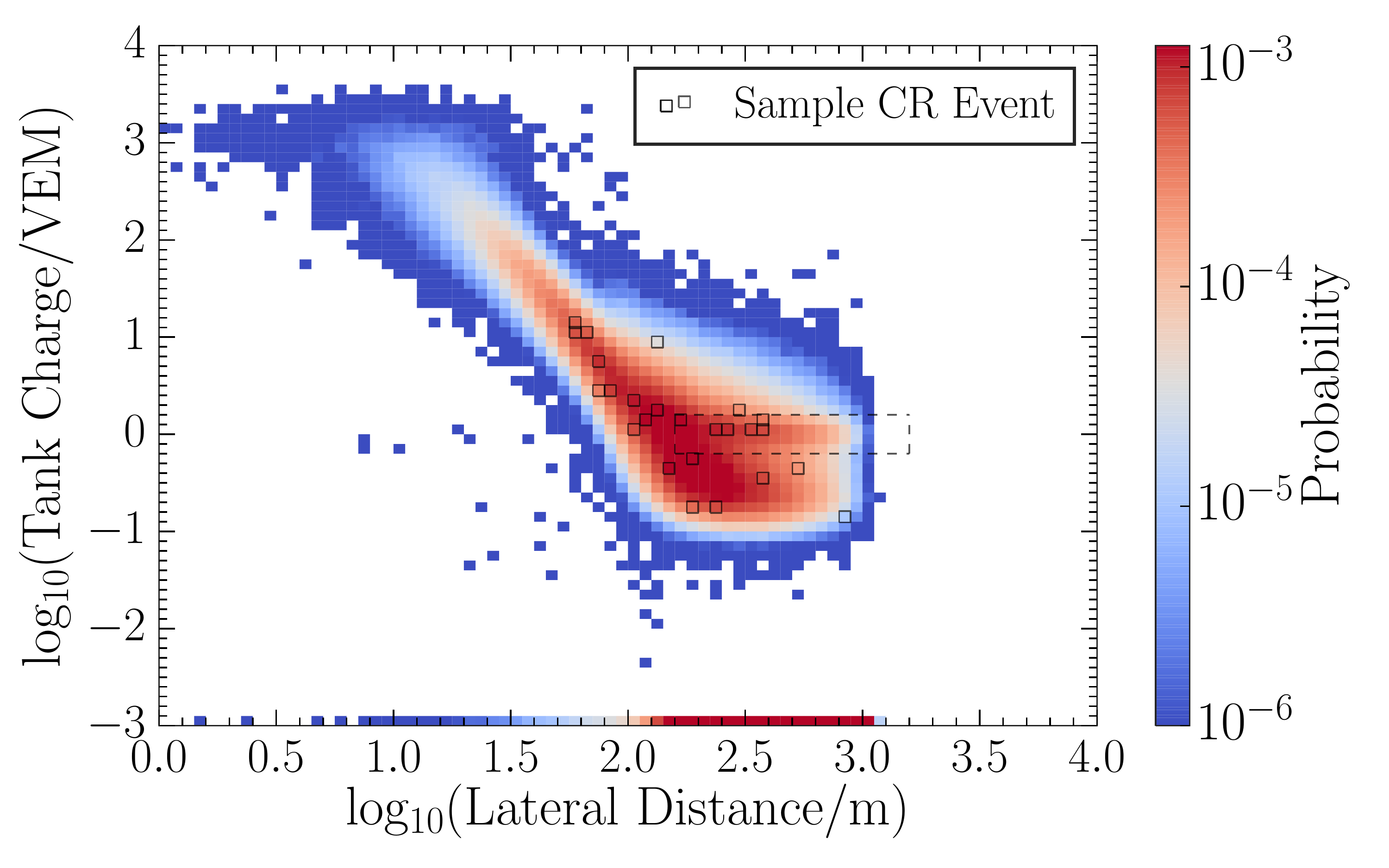}
\caption{IceTop PDFs, describing the charge distribution as function of the lateral distance of the tank from the shower axis, for simulated gamma-ray (left) and observed cosmic-ray (right) events with 0.3$\,\leq\,$log$_{10}$(S$_{125}$)$\,<\,$0.4 and 0.95$\,\leq\,$cos($\theta$)$\,<\,$1.0. Hit tanks for one sample cosmic-ray event are indicated by open boxes on both PDFs and the feature attributed to GeV muons is highlighted using dashed lines.}
\label{fig:thumb_plots}
\end{figure*}

\subsubsection{IceTop Shower Footprint}
\label{sec:llhratio}
As discussed in Section \ref{sec:detector}, minimum ionizing muons passing through an IceTop tank deposit charge such that the peak of the muon charge distribution is at $\sim$1~VEM with a width attributed to the zenith angle distribution of muons~\citep{Abbasi:2013}.  At sufficiently large distance from the shower core, the electromagnetic component becomes sub-dominant and the characteristic $\sim$1~VEM signals from GeV muons can be discerned. Figure \ref{fig:thumb_plots} shows a probability distribution function (PDF) describing the distribution of the charges as function of the lateral distance from the reconstructed shower core, also called LDF, for simulated gamma rays and observed cosmic rays. The prominent muon signal can be seen emerging in the cosmic-ray PDF beyond $\sim$200~m while it is very diminished in the gamma-ray PDF. The local charge fluctuations, observed as the width of the charge distribution for a given lateral distance in Figure \ref{fig:thumb_plots}, is also a measure of the hadronic content of the shower. The longitudinal development stage of the shower is reflected in the slope of the LDF seen in Figure \ref{fig:thumb_plots}. The curvature of the shower front, i.e. arrival time distribution of particles as a function of the lateral distance, is also sensitive to the longitudinal stage of the shower as it reaches the IceTop surface.

We construct three two-dimensional PDFs that incorporate these shower front properties. For this we use information from individual IceTop tanks, which are indexed from 1$\,\leq\,$i$\,\leq\,$162.  The PDFs are constructed using tank charges $\{Q_{\mathrm{i}}\}$, their lateral distances from the reconstructed shower axis $\{R_{\mathrm{i}}\}$, and hit times with respect to the expected planar shower front arrival time $\{\Delta T_{\mathrm{i}}\}$. Gamma-ray simulations and 10\% of cosmic-ray data are used to construct the PDFs for the gamma-ray $\{H_\gamma\}$ and cosmic-ray $\{H_{CR}\}$ hypotheses, respectively. Unhit and inactive tanks are included in the PDFs by assigning artificial and fixed values to charge and time ($Q_{\mathrm{i}}$=0.001~VEM and $\Delta T_{\mathrm{i}}$=0.01~ns) outside the range of hit tank values. The lateral distance distribution of unhit tanks (as seen in the bottom of plots in Figure \ref{fig:thumb_plots}) also contributes to the differences between the gamma-ray and cosmic-ray PDFs.

Based on each of the three two-dimensional PDFs, a log-likelihood ratio is calculated for all events. For instance, the log-likelihood ratio using the lateral charge distribution for a given event is defined as%
\begin{equation}
\Lambda_{QR} = \log_{10}\left(
						\frac{L_{QR}(\mathrm{event}| H_\gamma )}
                        {L_{QR}(\mathrm{event}| H_{CR})}
                        \right),
\end{equation}%
where the likelihood $L_{QR}$ is defined as%
\begin{equation}
 L_{QR}(\mathrm{event}| H ) = \prod_{\mathrm{i}=1}^{162} P\left(Q_{\mathrm{i}},\ R_{\mathrm{i}} \right | H),
\end{equation}%
with $P\left(Q_{\mathrm{i}},\ R_{\mathrm{i}} \right | H)$ being the probability of observing a tank with measured charge $Q_{\mathrm{i}}$ and at lateral distance $R_{\mathrm{i}}$, for the hypothesis $H$. Hit tanks for a sample cosmic-ray event, overlaid on PDFs in Figure \ref{fig:thumb_plots} using hollow boxes, illustrate how such an event would collect a greater likelihood from the cosmic-ray PDF as compared to the gamma-ray PDF.

Similarly, one can calculate $\Lambda_{Q \Delta T}$ and $\Lambda_{\Delta TR}$ from the PDFs that describe the time distribution of charges and the shower front curvature. The sum of all three log-likelihood ratios is then used as an input to a random forest classifier described in Section \ref{sec:rf}. The hadronic content and the longitudinal development stage of the shower at the surface depend on the primary energy and zenith angle in addition to mass of the primary particle. To reduce this dependence, the construction of PDFs and calculation of the log-likelihood ratio is carried out in $\log_{10}(S_{125})$ bins of 0.1 and $\cos(\theta)$ bins of 0.05.

\subsubsection{Random Forest}
\label{sec:rf}

The final event selection is performed using a random forest classifier implemented using the open-source python software Scikit-learn~\citep{Pedregosa:2011}.  In total, five features are included in the training process:%
\begin{enumerate}
\item The total charge deposited in the in-ice array after applying the cleaning procedure described in Section 3.1.1.
\item The likelihood sum as described in Section 3.1.2.
\item The energy proxy $\log_{10}(S_{125})$.
\item The cosine of the reconstructed zenith angle.
\item A parameter which describes the containment of the shower axis within the in-ice array, defined to be $C = R/r$ , where $R$ and $r$ are defined in Figure~\ref{fig:containment}.
\end{enumerate}




To prevent over-training, the maximum tree depth is limited to 8.  The output of the random forest is a score between 0 and 1, with 1 being the most gamma-ray-like.  The signal threshold for classifying events as gamma rays is chosen to be 0.7.  These values were optimized through a cross-validation grid search on sensitivity performance.  Tuning the additional hyper-parameters of the random forest showed negligible impact and they were kept at their default values.

\begin{figure}[htb!]
\plotone{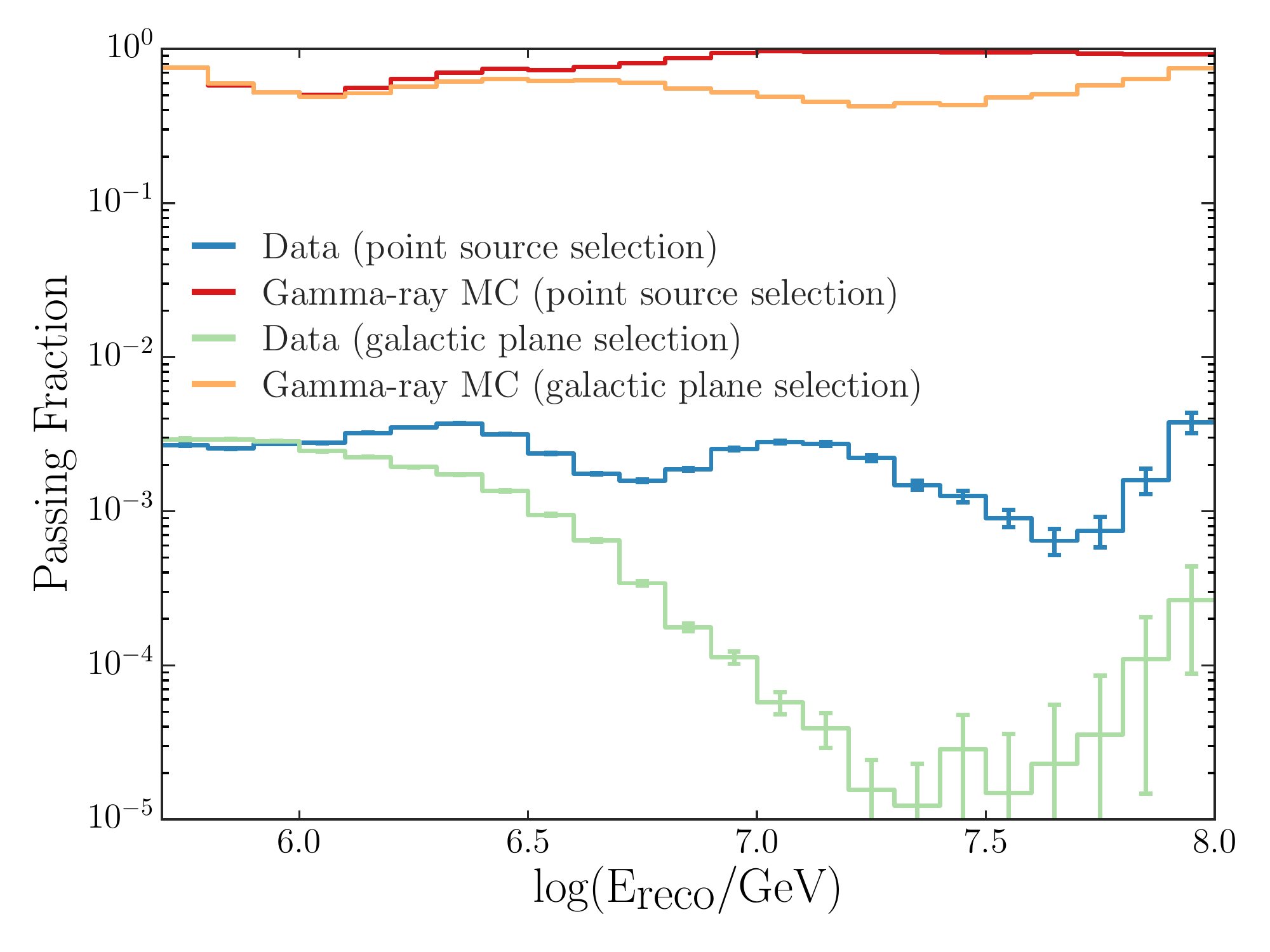}
\caption{Fraction of events 
which pass the gamma-hadron discrimination cut for gamma-ray simulation and data (cosmic-ray background) as a function of energy.  Both the point source and Galactic plane component event selections are shown.}
\label{fig:passing_fraction}
\end{figure}

Due to the differences in flux expectation, the gamma-ray spectrum used for training the classifier is different for the point source and diffuse cases.  For point sources, in order to be robust in performance to a range of spectral indices, we use two classifiers - one trained with a relatively hard (E$^{-2.0}$) spectrum and the other with a relatively soft (E$^{-2.7}$)  spectrum. In this case, the event is retained if either classifier returns a score above the signal threshold value.  For the diffuse case a single random forest trained with an E$^{-3.0}$ spectrum is used (see Section \ref{sec:gpresult} for the motivation behind the choice of spectral index).  Figure~\ref{fig:passing_fraction} illustrates the fraction of events that pass the signal threshold cut as function of energy.  The total number of events classified as gamma rays under this criteria for both cases are listed in Table~\ref{table:event_numbers_table} for each analysis year.

\section{Likelihood Analysis Methods}

All source hypotheses considered in this analysis were tested through an unbinned likelihood ratio method following the prescription of~\citet{Braun:2008}.  The form of the likelihood is dependent on the source class considered.

\subsection{Point Sources}
\label{sec:psllhmethod}

Sources that are point-like or extended in TeV gamma-ray astronomy should both appear point-like in this analysis. Hence, we construct a point source hypothesis for an unbiased source search in our entire field of view as well as for targeted H.E.S.S. source searches. The likelihood under this assumption takes the form: 
\begin{equation}
\label{eq:unbinned}
\begin{split}
L=\prod_j\;\prod_{i\in j}\left(\frac{n_s^j}{N^j}{S_i^j}\left(\left|\mathbf{x}_i-\mathbf{x}_S\right|,E_i,\sigma_i;\gamma\right) \right. \\
\left. +\left(1-\frac{n_s^j}{N^j}\right){B_i^j}\left(\delta_i,E_i\right)\right)
\end{split}
\end{equation}%
The likelihood $L$ is a product over \textit{$i$} events in each of \textit{$j$} datasets, where each dataset is comprised of one year of data. For a dataset $j$, $n_s^j$ is the number of signal events originating from the point source and $N^j$ is the total number of events.  Each event has a direction $\mathbf{x}_i = (\alpha_i, \delta_i)$ consisting of a right ascension $\alpha_i$ and declination $\delta_i$, an
energy $E_i$, and an angular uncertainty $\sigma_i$.  The events are compared to a point-source hypothesis comprised of a direction $\mathbf{x}_S$ and spectral index $\gamma$.  For the single source case the signal PDF $S^j_i$ is defined as:%
\begin{equation}
\label{eq:sig_pdf}
S_i^j=\frac{1}{2\pi\sigma^2_i}{\rm e}^{-\frac{\left|\mathbf{x}_i-\mathbf{x}_{S}\right|^2}{2\sigma_i^2}}\mathcal{E}_{S,i}^j\left(E_i, \delta_i, \gamma\right)
\end{equation}
where the angular uncertainty is included using a Gaussian distribution with a $\sigma_i$.
 Here, $\mathcal{E}^j_{S,i}$ is the normalized signal energy distribution.  The background PDF is defined as:%
\begin{equation}
\label{eq:bg_pdf}
B_i^j=\frac{1}{2\pi}B_{\rm exp}^j\left(\delta_i\right)\mathcal{E}_{B,i}^j\left(E_i, \delta_i\right)
\end{equation}
where $B_{exp}^j$ is the declination-dependent detector acceptance to cosmic rays derived from data, and $\mathcal{E}_{B,i}^j$ is the normalized background energy distribution.  The background PDF is uniform in right ascension and constructed from cosmic-ray data randomized in right ascension.

The likelihood is maximized with respect to $n_s$ and $\gamma$, where $n_s$ is the total number of signal events distributed among the signal events $n_s^j$ of each dataset proportionally according to the effective area and livetime of the samples.  This yields best fit values $\hat{n}_s$ and $\hat{\gamma}$ and a test statistic defined as:%
\begin{equation}
TS = -2\log\left(
				 \frac{L(n_s=0)}
                 {L(\hat{n}_s, \hat{\gamma})}
                 \right).
\end{equation}%
To evaluate the significance of an observed test statistic, a background test statistic ensemble is constructed from scrambled data using random right ascension values for the event directions.  The p-value is the fraction of the ensemble that have a test statistic exceeding the observed value.  When relevant, the number of independent trials performed (e.g. the number of H.E.S.S. source locations tested individually) is accounted for and a post-trial p-value is reported for the search.

To gauge the analysis sensitivity to point sources, a range of simulated fluxes are injected on top of scrambled data events.  The sensitivity is defined as the flux which produces a test statistic above the median of the background-only trial ensemble at the injected direction in 90$\%$ of the trials.  This is equivalent to the Neyman 90$\%$ confidence level construction~\citep{Neyman:1941}.  The discovery potential is defined as the flux which achieves a 5$\sigma$ detection in 50$\%$ of the trials.

When searching for emission from the selected H.E.S.S. point sources, we include a test for signal from all sources combined.  This stacking approach requires a modification to the likelihood, which we implement following the method in~\citet{Aartsen:2016b}.  For a catalog of $M$ point source locations, the signal PDF is constructed as:%
\begin{equation}
\label{eq:stacking_sig_pdf}
S_i^j = \frac{\sum_{m}^{M}\frac{R^j(\delta_m)}{2\pi\sigma^2_i}{\rm e}^{-\frac{\left|\mathbf{x}_i-\mathbf{x}_{S}\right|^2}{2\sigma_i^2}}}{\sum_{m}^{M}R^j(\delta_m)}\mathcal{E}_{S,i}^j\left(E_i, \delta_i, \gamma\right)
\end{equation}%
where $R_m^j$ is the relative detector acceptance to gamma rays at the location of the source $m$.  In this form, the sources are weighted assuming equal flux at Earth for each source.

\subsection{Diffuse Source}
\label{sec:diffusellhmethod}

The source hypotheses for the Galactic plane and cascade neutrino (see Section~\ref{sec:heseresult}) searches extend spatially over a significant portion of the sky.  For these cases, it is no longer valid to treat the signal as having a negligible contribution to the background PDF by averaging the right ascension.  Instead, a modification to the likelihood is made, following a method first introduced by \citet{Aartsen:2015} in a binned likelihood approach and later applied in an unbinned likelihood by \citet{Aartsen:2017b}, whose formulation we use here.

The background term $B_i^j$ in Equation~\ref{eq:unbinned} is replaced with two terms, $\tilde{D}_i^j$ and $\tilde{S}_i^j$, which are the event densities of the experimental data and gamma-ray simulation, respectively, after averaging over right-ascension:%
\begin{equation}
\label{eq:diffuse_llh}
\begin{split}
L = \prod_j\;\prod_{i\in j}\left( \frac{n_s^j}{N^j} S_i^j(\mathbf{x}_i, \sigma_i, E_i; \gamma) + \tilde{D}_i^j(\sin \delta_i, E_i) \right.\\
\left. - \frac{n_s^j}{N^j} \tilde{S}_i^j(\sin \delta_i, E_i) \right)
\end{split}
\end{equation}%

The construction of the signal PDF $S$ begins with a model of the gamma-ray flux distribution over the entire field of view.  This raw signal term must be convolved with the detector response to produce the expected observed distribution of emission in direction and energy. This is executed through a bin-by-bin multiplication of the relative detector acceptance to gamma rays, determined through simulation, and the flux model.  The point spread function of each event, described as a Gaussian distribution of width $\sigma$, is accounted for through the convolution of the signal map with a range of $\sigma$ from 0.1$^{\circ}$ to 1.0$^{\circ}$ in steps of 0.05$^{\circ}$.  On an event-by-event basis, the map corresponding to the $\sigma$ of the event is used as the signal PDF $S$.

\section{Results}

\subsection{All-Sky Point Source Search}
\label{sec:psresult}

\begin{figure*}
\plotone{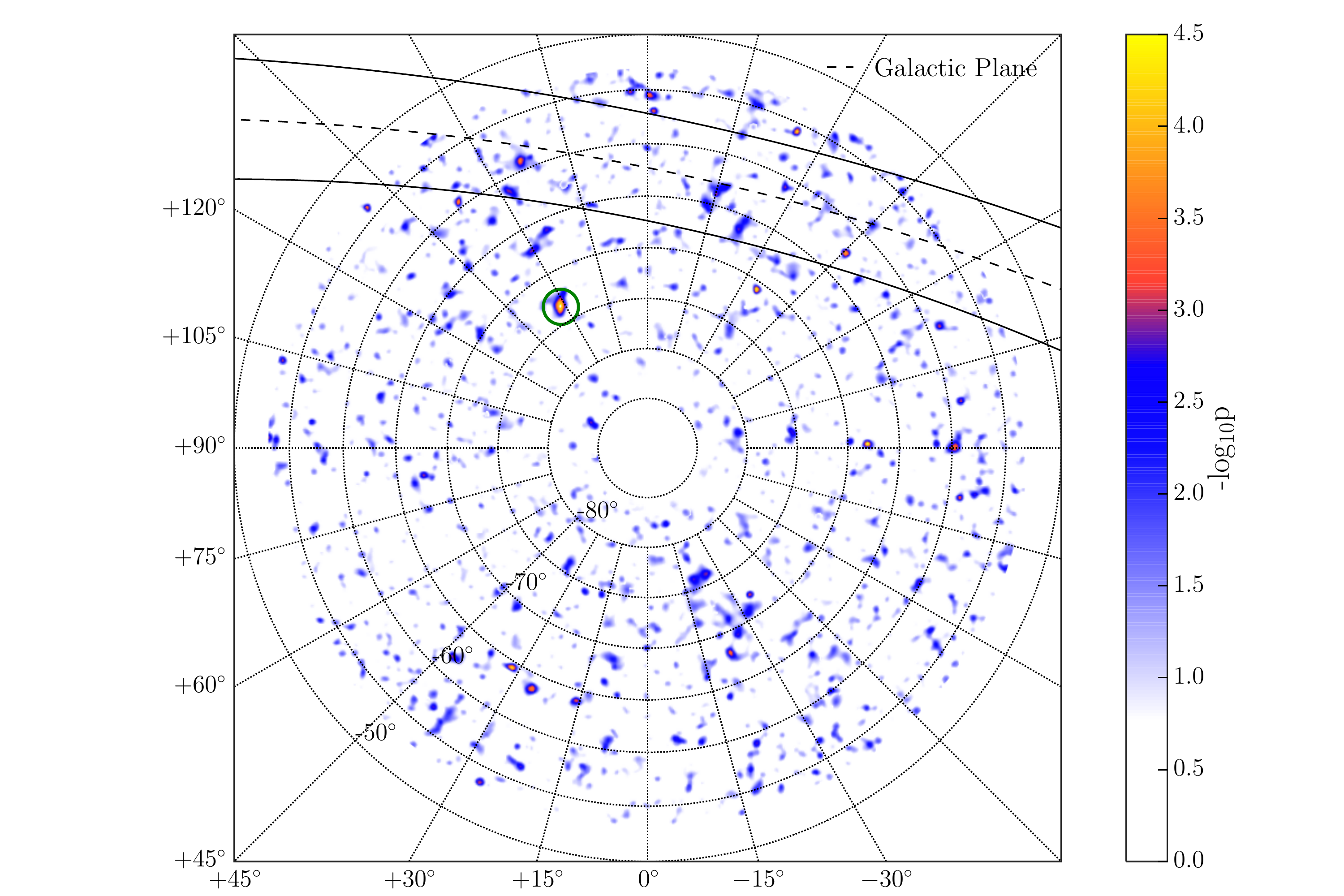}
\caption{All-sky likelihood scan pre-trial p-values shown projected from the South Pole in equatorial units.  The right ascension is labeled along the figure axes, with the interior text denoting declination bands.  The green circle highlights the hottest spot in the scan.  The Galactic plane region ($<$5$^{\circ}$ in Galactic latitude) is also shown.}
\label{fig:all_sky_scan}
\end{figure*}

An unbiased search for a point source is accomplished by scanning over the entire field of view.  The position of a single point source is assumed to lie in the direction of a given pixel in a HEALPIX map (N$_{side}$=512, pixel diameter of 0.11$^{\circ}$)~\citep{Gorski:2005}.  A test statistic is calculated using Equation~\ref{eq:unbinned} under this hypothesis.  This process is repeated for each pixel in the analysis field of view.  The resulting p-values of this scan, before accounting for trials, are show in Figure~\ref{fig:all_sky_scan}.  The region of the sky with zenith angle $<$5$^{\circ}$ is excluded, as within this region scrambling in right ascension alone is insufficient to build independent background trials.

\begin{figure}[htb!]
\plotone{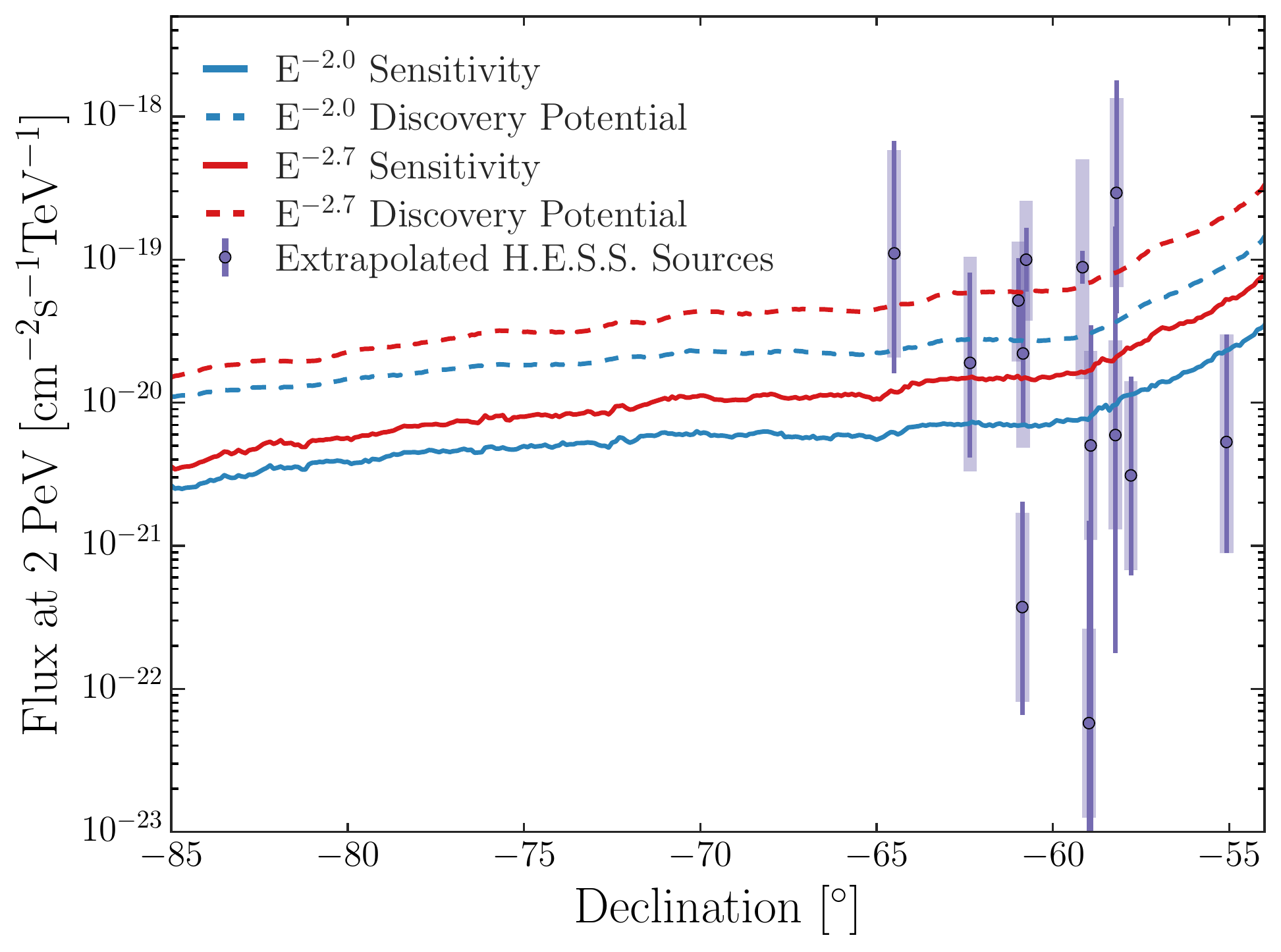}
\caption{The sensitivity and discovery potential thresholds to a gamma-ray flux at Earth for E$^{-2.0}$ (solid) and E$^{-2.7}$ (dashed) point sources at 2 PeV as a function of declination are shown in blue and red, respectively.  In purple are the extrapolated fluxes up to 2 PeV of H.E.S.S. sources in the analysis field of view, under the scenario of no break in the best-fit energy spectrum, with an approximate correction for absorption using model predictions from \citealt{Lipari:2018}.  Error bars indicate the statistical uncertainty, while the systematic uncertainty is represented by the shaded boxes.}
\label{fig:ps_sensitivity}
\end{figure}

The hottest spot in the sky, with a pre-trial p-value of \num{4e-5}, is located at -73.4$^{\circ}$ in declination and 148.4$^{\circ}$ in right ascension, with $n_s$ = 67.9$^{+17.8}_{-16.6}$ and a spectral index of 2.9$^{+0.3}_{-0.3}$.  The post-trial p-value, calculated by comparing the observed test statistic to the background ensemble of hottest-spot test statistic values, is 0.18, consistent with background expectation.  Figure~\ref{fig:ps_sensitivity} shows the sensitivity and discovery potential to point sources of this analysis as a function of declination.  The proportion of showers which are contained in IceCube drops sharply at higher shower inclinations, resulting in a decrease in performance at high declination values.

\input{hess_table.tex}

\subsection{TeV Gamma-Ray Source Studies at PeV Energies}
\label{sec:hessresult}
\begin{figure*}[htb!]
\plotone{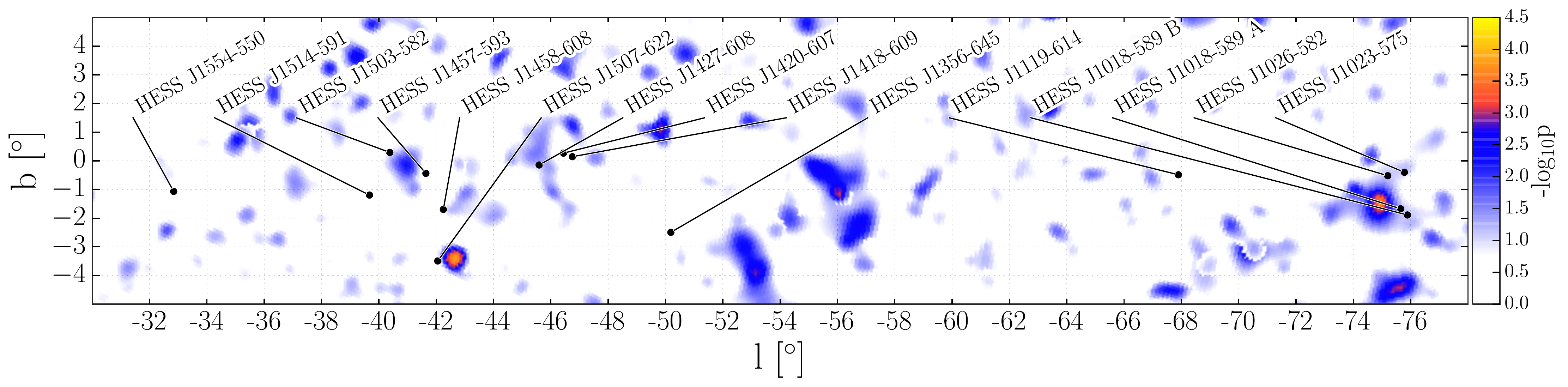}
\caption{All-sky likelihood scan pre-trial p-value shown in Galactic coordinates for $|$b$|$ $<$ 5.  The H.E.S.S. sources in the analysis field of view are shown in black.}
\label{fig:hess_sources}
\end{figure*}

There are a total of 15 TeV gamma-ray sources in the field of view of this analysis which have no evidence of a cut-off in their energy spectrum; all of these sources were reported by the H.E.S.S. collaboration.  Table~\ref{table:hess_p_values} lists the name of each source paired with the citation which provided the H.E.S.S. fit information, along with the best-fit declination and spectral index observed by H.E.S.S..  The projected flux at 2 PeV of these sources assuming there is no cut-off in their energy spectra are shown in Figure~\ref{fig:ps_sensitivity}.  Attenuation effects were calculated for the galactic coordinates and distance of each source using model results provided by~\citet{Lipari:2018}.  Source distances were estimated from associated x-ray or radio observations when possible~\citep{Wakely:2008}.  Otherwise, a distance of 8.5 kpc, roughly that of the Galactic center, is assumed.  The directions of the sources are shown overlaid on the all-sky scan in Figure~\ref{fig:hess_sources}.

In the first test of these sources, each source location is evaluated independently, making no assumption on the spectral index of the source.  The pre-trial p-value for every considered source resulting from this test is reported in Table~\ref{table:hess_p_values}.  The most significant individual source, HESS J1427-608, has a pre-trial p-value of 0.07, with a fitted $n_s$ of 25.9$^{+16.6}_{-15.7}$ and spectral index of 3.2$^{+0.8}_{-0.7}$.  This results in a trials-corrected p-value of 0.65, consistent with background expectation.

To place constraints on the fluxes of these sources individually, we consider the case where the flux of each source extends with no cut off in energy and with the spectral index observed by H.E.S.S.. Under this flux assumption, we report the 90\% confidence level upper limit on the flux at 2 PeV at Earth for each source in Table~\ref{table:hess_p_values} as $\Phi_{90\%}$. We note that uncertainties in the best-fit H.E.S.S. spectrum have a negligible effect on the upper limits and their relationship to the extrapolated flux.

For a subset of the sources our limits are strong enough to place constraints on an energy cut-off of the source flux.  Here we assume the flux to be the combination of a power law and an exponential energy cut-off at $E_{cut}$:
\begin{equation}
\label{eq:energy_cut_off}
\Phi(E) = \Phi_{\textrm{P.L.}}(E)e^{-E/E_{cut}}
\end{equation}
The value of $E_{cut}$  was determined by evaluating, for a range of energy cut-off values, the resulting flux of the source and the analysis sensitivity for the same hypothesis.  Here we do incorporate absorption from~\citet{Lipari:2018} in order to limit the energy cut off at the source.  The minimum energy cut-off values where our sensitivity lies below the flux expectation of the source is reported in Table~\ref{table:hess_p_values} as a 90\% confidence level upper limit on $E_{cut}$.

We additionally search for combined PeV emission from all the H.E.S.S. sources. In this test all the sources are stacked, using the modified signal PDF from Equation~\ref{eq:stacking_sig_pdf}, which weights the sources assuming equal flux at Earth for each source.  The result of this stacking correlation test is a p-value of 0.08, consistent with background.

\subsection{IceCube HESE Neutrino Correlation}
\label{sec:heseresult}

\begin{figure*}
\plotone{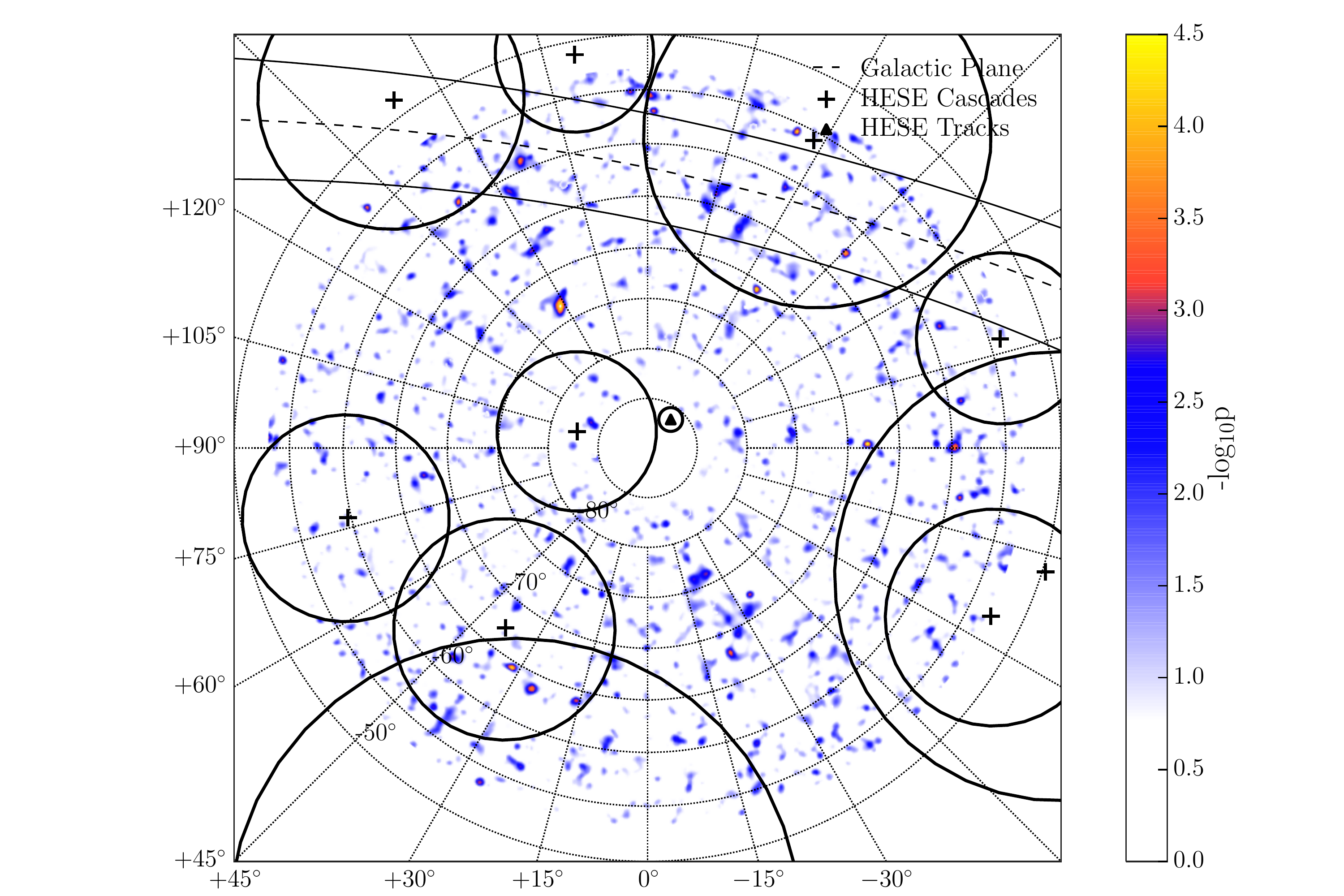}
\caption{The reconstructed directions and 1$\sigma$ uncertainties of events from the four-year HESE neutrino sample superimposed on the all-sky-scan results.}
\label{fig:HESE_events}
\end{figure*}

From the four year high-energy starting event (HESE) neutrino sample~\citep{Aartsen:2015b} a total of eleven events lie within the field of view of this analysis. The event positions and 1$\sigma$ uncertainties are overlaid on top of the pre-trial p-value map of the all-sky point-source search in a polar projection in Figure~\ref{fig:HESE_events}.  There are two topologically distinct event types, referred to as cascades and tracks.  We treat the event types separately.

Cascade neutrino events are produced by neutral-current interactions as well as charged-current interactions of electron and tau neutrinos.  These events have poor angular resolution, typically greater than 10$^{\circ}$ in the HESE sample.  In order to correctly account for the change in detector acceptance over such a large point spread function, we test for a correlation with these cascade neutrino events using the template likelihood method described in Section 4.2. The signal template is constructed by combining the spatial likelihood contour PDFs for each event.  The correlation test resulted in a conservative p-value of $>\,$0.5.  We place a 90$\%$ confidence level upper limit of \num{1.07e-19}~\si{cm^{-2}s^{-1}TeV^{-1}} on the flux at 2 PeV of a source class consistent with the HESE cascade directions and an E$^{-2.0}$ spectrum.

Track neutrino events are the product of charged current muon neutrino interactions. These interactions produce muons that can travel several kilometers through the ice, which allows for reconstructions with angular resolution $<\,$1.2$^{\circ}$ for events in the HESE sample.  There is one track event in our field of view, which we treat under the point source prescription.  At a declination of $\delta$=-86.77$^{\circ}$, scrambling events in right ascension alone is insufficient to build a background test statistic distribution; instead, we scramble events randomly throughout the polar cap region of $<$5$^{\circ}$ zenith angle, within which acceptance was found to be approximately even.  No evidence of signal was found in test for a point source at the event location, which resulted in a conservative p-value of $>$0.5.

\begin{figure*}[htb!]
\centering
\includegraphics[height=5.5cm]{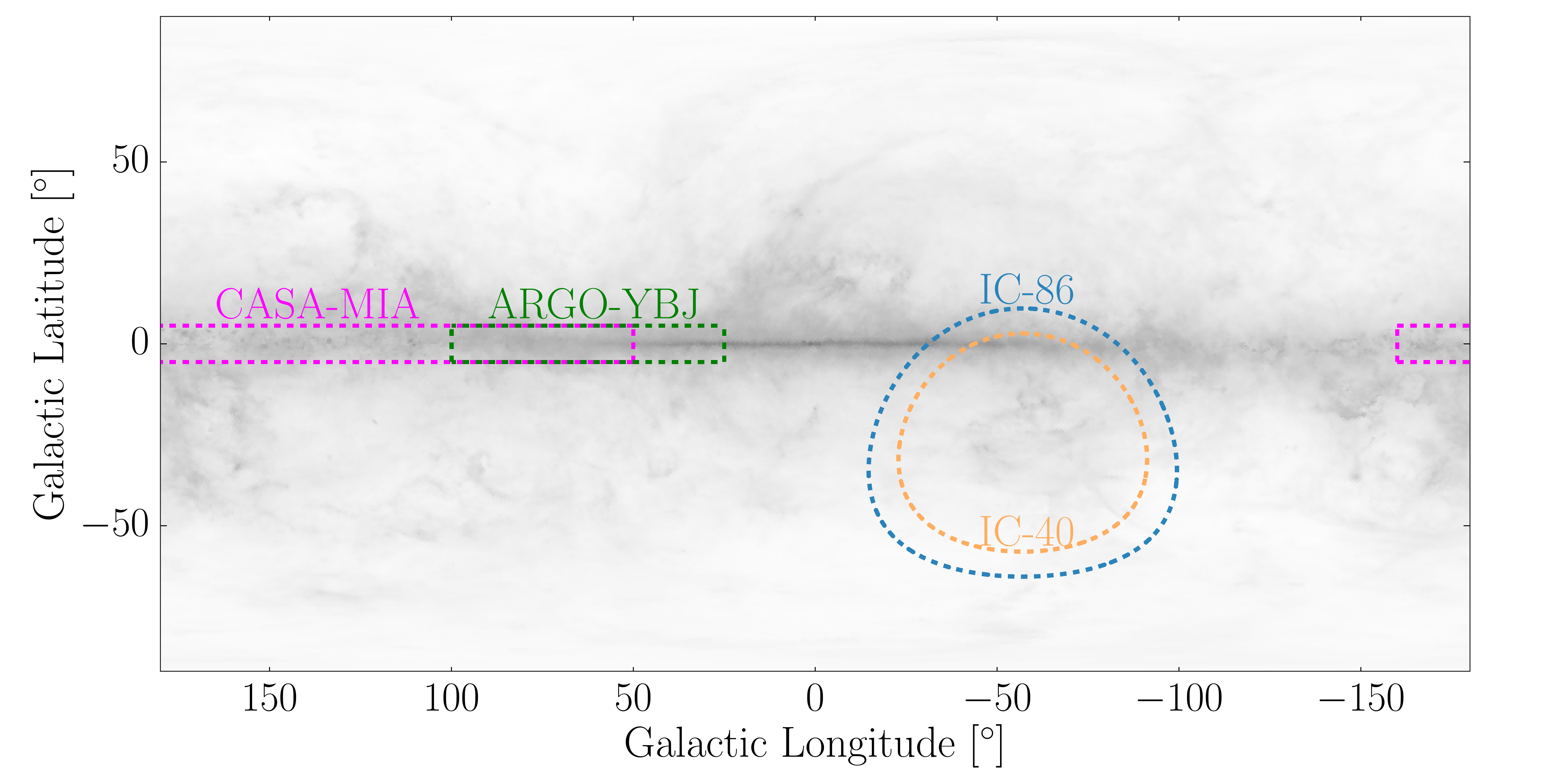}\hspace{0.3cm}\includegraphics[height=5.5cm]{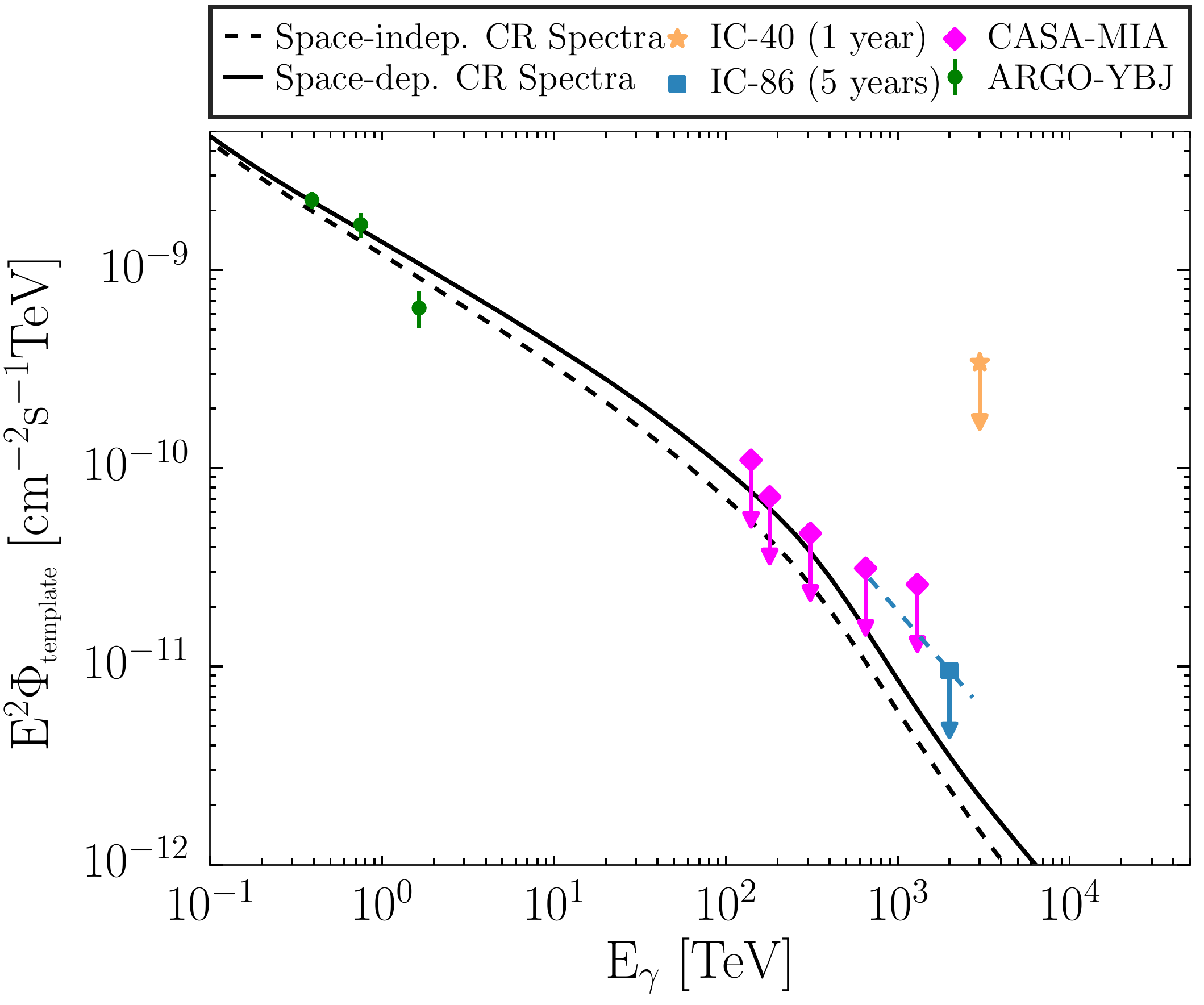}
\caption{Left:  The respective field of views of CASA-MIA~\citep{Borione:1998}, ARGO-YBJ~\citep{Bartoli:2015}, IC-40~\citep{Aartsen:2013}, and this analysis overlaid on a map of the $\pi^0$ decay component of the Fermi-LAT Galactic plane diffuse emission model~\citep{Ackermann:2012}.  Right:  The IceCube 90$\%$ confidence level upper limit (IC-86) on the angular-integrated scaled flux from the Galactic plane in our field of view for a spatial distribution of emission given by the $\pi^0$ decay component of the Fermi-LAT diffuse emission model. The IC-86 upper limit is compared to results from ARGO-YBJ, CASA-MIA, and IC-40, using the scaling defined in Equation~\ref{eq:scaledflux}. Dotted lines show the E$^{-3}$ spectrum, used for obtaining IceCube upper limits, over the energy range containing 5$\%$ to 95$\%$ events in the final sample. Also shown are, for two models, absorbed flux predictions for the IC-86 field of view calculated by the authors of ~\cite{Lipari:2018} on special request. The two models assume space-independent and space-dependent cosmic-ray spectra throughout the Galaxy, respectively.}
\label{fig:fermi_limit}
\end{figure*}

\subsection{Diffuse Galactic Plane Template Analysis}
\label{sec:gpresult}

\begin{figure*}
\plotone{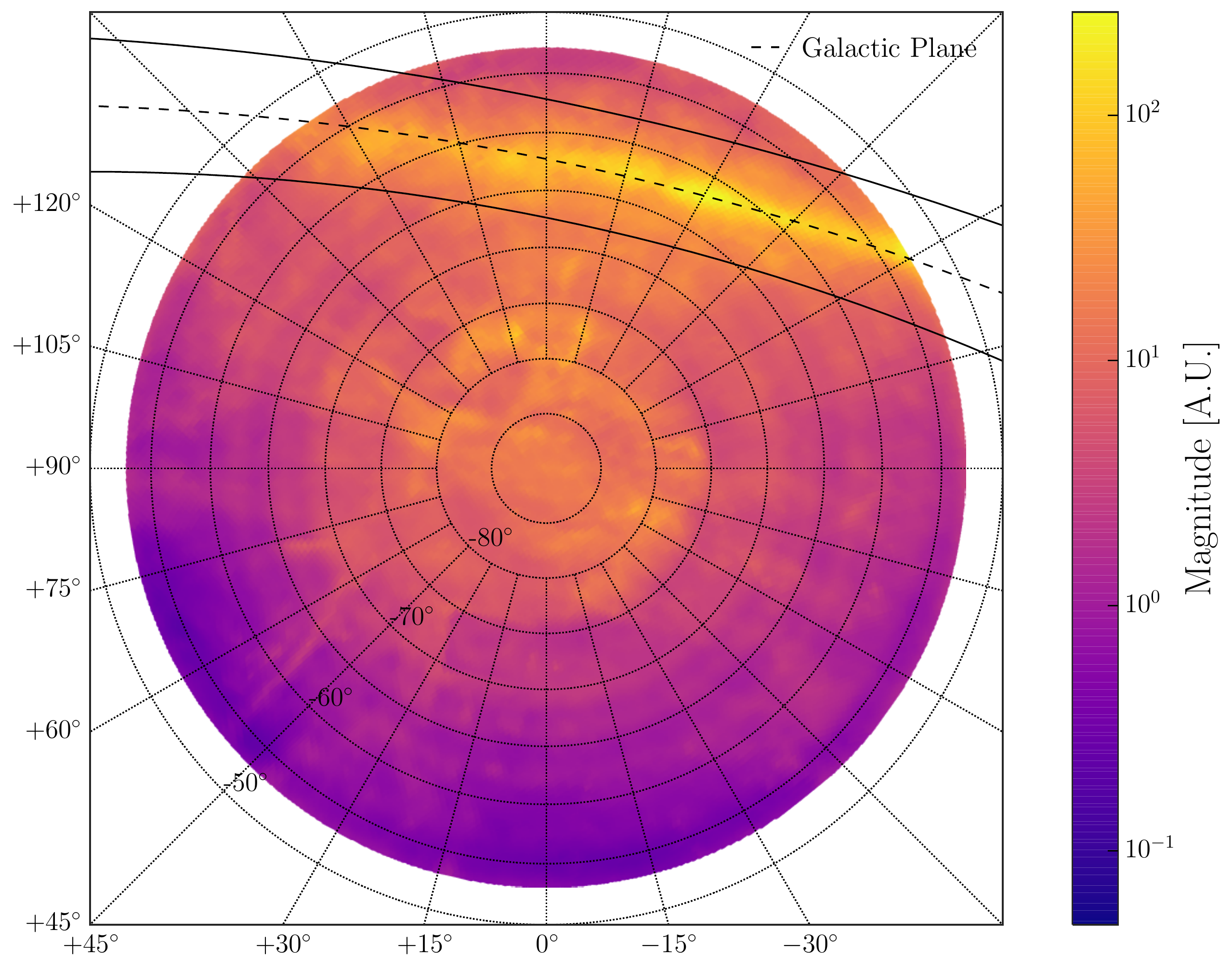}
\caption{The Fermi-LAT $\pi^0$ decay spatial template multiplied by the detector acceptance for the data year 2012.}
\label{fig:fermi_pi0_template}
\end{figure*}

To test for a diffuse flux from the Galactic plane, the template likelihood method is employed with the signal template taken to be the pion decay component of the Fermi-LAT diffuse emission model~\citep{Ackermann:2012}.  The Fermi-LAT template multiplied by the detector acceptance is shown in Figure~\ref{fig:fermi_pi0_template} for the 2012 sample.

The expected observed spectral index of diffuse gamma rays from the Galactic plane is still an open question due to existing uncertainties in the Galactic cosmic-ray spectrum, interstellar gas distribution, and flux attenuation.  The unattenuated flux at PeV energies has been predicted to be as hard as E$^{-3}$ (\citealt{Vernetto:2017}, \citealt{Ingelman:1996}), although the spectrum could be as soft as E$^{-3.4}$~\citep{Ingelman:1996} after attenuation.  Here we fix the spectral index to 3.

Maximization of the likelihood in Equation \ref{eq:diffuse_llh} returns an observed test statistic corresponding to a p-value of 0.28, which provides no evidence of a diffuse signal from the Galactic plane under the current hypothesis. We place a 90$\%$ confidence level upper limit of \num{2.61e-19}~\si{cm^{-2}s^{-1}TeV^{-1}} on the normalization of the spectral energy distribution at 2~PeV with spectral index 3.  The normalization energy was chosen to be 2~PeV as that is the energy for which the analysis was found to be least sensitive to the spectral index assumption.

As previous experimental limits have used a box region around the Galactic plane to describe the diffuse emission, for a limit comparison we use an angular-integrated scaled flux%
\begin{equation}
\Phi_{\mathrm{template}} = \Phi \Delta\Omega \, \frac{\int_{\mathrm{all-sky}} \mathrm{S}_{\mathrm{Fermi}} d\Omega }{ \int_{\Delta \Omega} \mathrm{S}_{\mathrm{Fermi}} d\Omega },
\label{eq:scaledflux}
\end{equation}%
where the angular-integrated flux from the observed region is $\Phi \Delta \Omega$, and the second term scales this flux by the fraction of the Galactic plane emission present in the observed region as given by the Fermi-LAT pion decay template S$_{\mathrm{Fermi}}$.  This is, in effect, a limit on the overall normalization of the Fermi template flux.  Figure \ref{fig:fermi_limit} compares the scaled flux limits from this analysis with the existing IC-40~\citep{Aartsen:2013} and CASA-MIA~\citep{Borione:1998} limits, as well as the measured flux by ARGO-YBJ~\citep{Bartoli:2015}.  In addition, the gamma-ray flux from the diffuse Galactic plane in the analysis field of view is shown for two model predictions from~\cite{Lipari:2018}.  The first assumes that cosmic-ray spectra have a space independent spectral shape throughout the Galaxy.  The second assumes the central part of the Galaxy has a harder cosmic-ray spectrum than observed at Earth, an idea supported by some recent analyses (\citealt{Gaggero:2015b}, ~\citealt{Yang:2016diffuse}).

\subsection{Systematics}\label{sec:systematics}
There are a number of systematic uncertainties that can affect the estimated sensitivity of the study, including the snow attenuation, the calibration of the charge deposited in IceTop tanks, the hadronic interaction model and the ice model used in simulations, and the anisotropy of the cosmic-ray flux.  For each systematic component, datasets are constructed using the parameter uncertainty bounds.  These datasets are propagated through the entire analysis chain to evaluate the corresponding uncertainty in sensitivity.  Here we report both point source and Galactic plane analysis sensitivity uncertainties.  For the point source sensitivity, we assume the median uncertainty based on a scan over the declination range in 0.1$^{\circ}$ bins.  These datasets were constructed from events not used in the training of the random forest classifier detailed in Section~\ref{sec:rf}.

The optimal value for the snow attenuation coefficient $\lambda$, used in the charge correction for IceTop tanks as shown in Equation~\ref{eq:snow_correction}, has been found to vary by $\pm$0.2 m over a range of zenith angle and energy values~\citep{Aartsen:2013c}.  A more thorough model of the snow attenuation is work in-progress; for this work we quote this bound as a conservative estimate.  This systematic results in an uncertainty of 11.0\% in sensitivity to point sources and 11.2\% in sensitivity to a diffuse flux from the Galactic plane.

The calibration of IceTop tank charge to VEM units requires a fitting of the muon peak in the charge spectrum of the tank~\citep{Abbasi:2013}.  The dependence of this fit to systematic factors was studied in detail by \citet{Van:2011}.  They found an uncertainty of at most $\pm$3\% on the charge calibration, which propagates directly to an uncertainty in the deposited signal.  This systematic error results in an uncertainty of 2.1\% in sensitivity to point sources and 7.4\% in sensitivity to a diffuse flux from the Galactic plane.

The number of muons generated in simulated gamma-ray air showers at energies sufficient to trigger the detectors is governed by the high-energy hadronic interaction model used in CORSIKA.  In order to evaluate the magnitude of the model dependence, we perform sensitivity studies with simulation generated using QGSJetII-04~\citep{Ostapchenko:2011}, but otherwise identical to the original set.
We chose QGSJetII-04 over other post-LHC models
since it was the model that produced the most muons in hadronic air showers~\citep{Plum:2018comp}. Sensitivities calculated with these systematic datasets resulted in a 23.2\% uncertainty for point sources and a 26.2\% uncertainty for a diffuse flux from the Galactic plane.

The anisotropy of the cosmic-ray flux is a potential source of signal contamination.  While declination-dependent anisotropy is accounted for due to the use of data to construct the background PDF in the likelihood, any anisotropy in right ascension is not.  However, within the analysis field of view this anisotropy is at a level of at most 0.03\%~\citep{Aartsen:2016}.  This is negligible in relation to statistical uncertainties, which is $\sim$25\% in flux at the sensitivity threshold.

In simulation, the uncertainties in the optical properties of the ice can affect the amount of charge measured.  While this a potential for systematic error, an analysis with datasets using $\pm$ 10\% in deposited charge in IceCube showed negligible impact on sensitivity compared to statistical fluctuations.  Finally, the method we use naturally corrects for any bias in the energy proxy. Any systematic biases in fitted n$_s$ and $\gamma$ values unaccounted for are found to be negligible when compared to statistical uncertainties.

Under the assumption that the errors discussed are independent and Gaussian distributed, the overall sensitivity uncertainty resulting from quadrature addition is 25.8\% for point sources and 29.4\% for the Galactic plane.

\section{Discussion and Conclusion}
We have presented the results of multiple searches for PeV gamma rays using five years of data from 2011 to 2016 collected by the IceCube Observatory.  For all flux hypotheses considered, no significant excess in emission above background expectation was observed.

An unbiased scan over the entire analysis field of view resulted in a declination-dependent 90\% confidence level upper limit of $\sim$10$^{-21}$ - 10$^{-20}$~\si{cm^{-2}s^{-1}TeV^{-1}} on the flux at 2 PeV of a gamma-ray point source, the most stringent PeV gamma-ray point source limits to date and an improvement of more than an order of magnitude over the previous IceCube analysis~\citep{Aartsen:2013}.

\begin{figure}[htb!]
\plotone{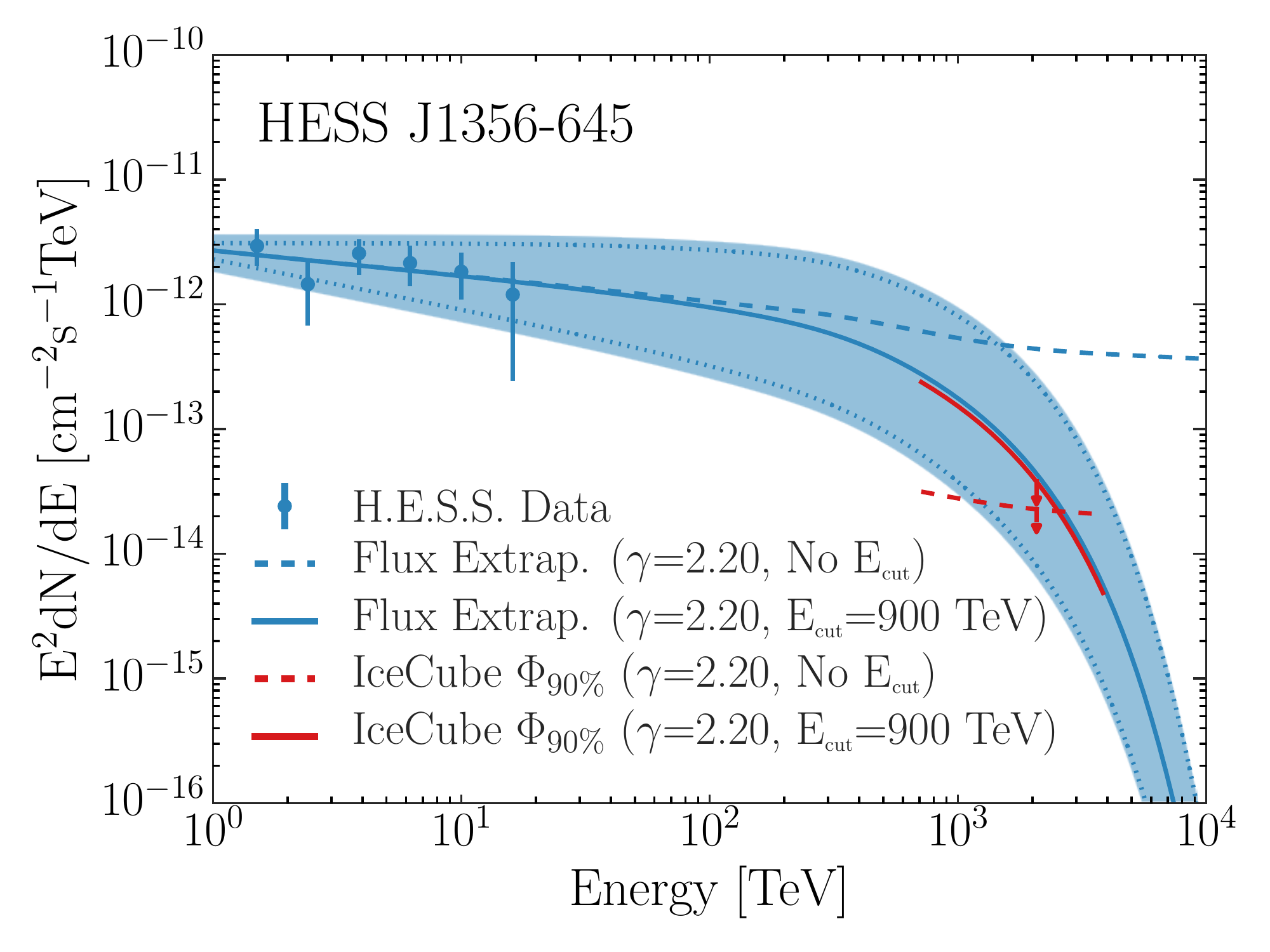}
\caption{Flux measurements of the source HESS J1356-645~\citep{Abramowski:2011}, along with a power-law spectra with (solid line) and without (dashed line) a 900 TeV cut off in energy, where for both absorption is included for the source distance of 2.4 kpc~\citep{Wakely:2008}.  For the cut off extrapolation, the shaded region denotes the statistical uncertainty, while the systematic uncertainty is represented by dotted lines.  The 90\% confidence level upper limit to these flux extrapolations are shown in red.}
\label{fig:hess_J1356}
\end{figure}

In addition, we searched for PeV gamma-ray emission of H.E.S.S. sources that show no evidence for a spectral cut off at TeV energies.  For the first time, upper limits have been placed on the high-energy emission of these sources.  For seven sources, these limits exclude the spectra observed by H.E.S.S. from extending without a cut off to PeV energies after accounting for absorption.  The lowest energy cut-off limit value is 900 TeV for the source HESS J1356-645.  Figure~\ref{fig:hess_J1356} illustrates the extrapolated flux for this source with a 900 TeV cut off in energy along with the corresponding 90\% upper limits set by this analysis to the same flux assumptions. Considering the systematic/statistical uncertainties in the H.E.S.S. measurements, in the measurement of the source distance, and those presented in Section~\ref{sec:systematics}, we do not expect sensitivity to spectral assumptions that slightly differ from our own. We present these constraints as order of magnitude estimates for other spectral assumptions which may have, for example, additional spectral softening or a cut off other than a simple exponential.

Since this analysis was executed, a new Galactic plane survey paper was published by the H.E.S.S. collaboration that included a reanalysis of the sources included in this study~\citep{Abdalla:2018}.  For all cases, the new analysis positions, spectral indices, and flux normalizations at 1 TeV are consistent with the values used for calculating the presented upper limits within statistical and systematic errors.  The largest discrepancy in fit values is for the source HESS J1427-608, which had a best-fit spectral index of 2.20 in \citet{Aharonian:2008}.  \cite{Guo:2017} reported on a counterpart seen in Fermi-LAT data at GeV energies with a best fit including the H.E.S.S. data from \citet{Aharonian:2008} of $E^{-2}$ over four orders of magnitude in energy with no break in the spectrum, a property unique among currently known TeV sources.  However, in ~\citet{Abdalla:2018} the reanalysis best-fit spectral index was found to be 2.85. 




The point-source map in the Galactic plane region shown in Figure \ref{fig:hess_sources} has several interesting spots with low p-values.  The first is spatially coincident with the binary source HESS J1302-638 at $b$ = -0.99$^{\circ}$ and $l$ = -55.81$^{\circ}$~\citep{Aharonian:2005}.  This source was excluded from the targeted search as TeV emission has only been observed during the periastron of the source, which occurs only once every 3.4 years due to a highly eccentric orbit~\citep{Romoli:2017}.  One such periastron occurred during the collection period of data used in this analysis in 2014.  However, a follow-up analysis with only the data from the 2014 period showed no evidence for PeV gamma-ray emission from this direction.  The second spot, near HESS~J1507-622, is interesting as both the source and the spot lie on the edge of a large, nearby CO molecular cloud observed by \citet{Dame:2001} which is most likely $\sim$400~pc away.  The type of the source has not been established, and it has a uniquely high Galactic latitude ($\sim$3.5$^{\circ}$)~\citep{Acero:2011}.  This allows for the possibility of the source being nearby.  However, the offset from the spot to the source location and the most dense region of the molecular cloud makes any association unlikely.

The correlation test between PeV gamma-ray candidate events and the neutrino events, from the IceCube HESE sample which lie within the field of view of this analysis, also yields a null result.

We place a 90$\%$ confidence level upper limit of \num{2.61e-19}~\si{cm^{-2}s^{-1}TeV^{-1}} on the angular-integrated diffuse gamma-ray flux from the Galactic plane at 2~PeV under the assumption of an E$^{-3}$ spectrum. Figure \ref{fig:fermi_limit} (left) compares the field of views of the different experiments. Here it is worth noting that the current analysis (IC-86) constrains PeV diffuse flux from a denser region of the Galactic plane than CASA-MIA~\citep{Borione:1998}. As shown in Figure \ref{fig:fermi_limit} (right), the IC-86 upper limit is an order of magnitude better than the IC-40 result, and the most stringent constraint on the diffuse flux above 1 PeV. The IC-86 upper limit is also compared to two model predictions for attenuated diffuse flux from the Galactic plane in the IceCube field of view as given by \citet{Lipari:2018}. Both models derive diffuse gamma-ray emission under appropriate assumptions for the cosmic-ray flux throughout the Milky Way, inter-stellar gas distribution, and gamma-ray absorption effects.
The model labeled as `space-dependent CR Spectra' includes a changing spectral index of the gamma-ray spectrum as a function of the distance to the Galactic center to reproduce the effects of a harder cosmic-ray spectrum near the Galactic center. The two model predictions are not too different for the IC-86 field of view since it observes a part of the Galactic plane sufficiently far from the Galactic center. Even though the IC-86 upper limit cannot constrain the partially-empirical model predictions by \citet{Lipari:2018}, it may serve as important data for other detailed simulations of Galactic cosmic-ray transport such as \cite{Gaggero:2015b}.



IceCube's sensitivity to the gamma-ray flux is expected to improve at a rate lower than the inverse square root of the livetime expected from additional exposure. This is due to the reduced acceptance to gamma-ray air showers with continued snow accumulation on IceTop tanks. A proposed scintillator array~\citep{Huber:2017} at the surface will improve the sensitivity to the electromagnetic shower component and counteract the degradation of the photon-sensitivity due to snow accumulation. Furthermore, radio antennas at the surface may improve the gamma-hadron separation and increase the sky coverage such that the Galactic Center comes into the field of view~\citep{Balagopal:2018}. In the long term, a 7.9~km$^3$ next generation IceCube detector~\citep{vanSanten:2017} is being designed along with a 75~km$^2$ surface scintillator array. Adding radio antennas and non-imaging air cherenkov telescopes to the surface array would provide sensitivity to PeV gamma-rays over a much larger field of view than the current detector. Recent hints of a PeVatron~\citep{hess:pevatron} near the Galactic center and the expected increase in the diffuse flux towards the Galactic center are strong motivators to search for PeV gamma rays in the future with higher sensitivity instruments.  A future gamma-ray analysis with the IceCube observatory would be complementary to the planned experiments LHAASO~\citep{LHAASO} and HiSCORE~\citep{hiscore} in the Northern Hemisphere.  Such an analysis would be aided in a search for galactic point sources by results from the proposed southern CTA site, which will provide even higher energy measurements than H.E.S.S. of possible PeVatrons in the Southern Hemisphere~\citep{cta:2013}.

\acknowledgments
This work is in memory of Stefan Westerhoff.  The IceCube collaboration acknowledges the significant contributions to this manuscript from Zachary Griffith and Hershal Pandya.  We acknowledge the support from the following agencies:
USA {\textendash} U.S. National Science Foundation-Office of Polar Programs,
U.S. National Science Foundation-Physics Division,
Wisconsin Alumni Research Foundation,
Center for High Throughput Computing (CHTC) at the University of Wisconsin-Madison,
Open Science Grid (OSG),
Extreme Science and Engineering Discovery Environment (XSEDE),
U.S. Department of Energy-National Energy Research Scientific Computing Center,
Particle astrophysics research computing center at the University of Maryland,
Institute for Cyber-Enabled Research at Michigan State University,
and Astroparticle physics computational facility at Marquette University;
Belgium {\textendash} Funds for Scientific Research (FRS-FNRS and FWO),
FWO Odysseus and Big Science programmes,
and Belgian Federal Science Policy Office (Belspo);
Germany {\textendash} Bundesministerium f{\"u}r Bildung und Forschung (BMBF),
Deutsche Forschungsgemeinschaft (DFG),
Helmholtz Alliance for Astroparticle Physics (HAP),
Initiative and Networking Fund of the Helmholtz Association,
Deutsches Elektronen Synchrotron (DESY),
and High Performance Computing cluster of the RWTH Aachen;
Sweden {\textendash} Swedish Research Council,
Swedish Polar Research Secretariat,
Swedish National Infrastructure for Computing (SNIC),
and Knut and Alice Wallenberg Foundation;
Australia {\textendash} Australian Research Council;
Canada {\textendash} Natural Sciences and Engineering Research Council of Canada,
Calcul Qu{\'e}bec, Compute Ontario, Canada Foundation for Innovation, WestGrid, and Compute Canada;
Denmark {\textendash} Villum Fonden, Danish National Research Foundation (DNRF), Carlsberg Foundation;
New Zealand {\textendash} Marsden Fund;
Japan {\textendash} Japan Society for Promotion of Science (JSPS)
and Institute for Global Prominent Research (IGPR) of Chiba University;
Korea {\textendash} National Research Foundation of Korea (NRF);
Switzerland {\textendash} Swiss National Science Foundation (SNSF);
United Kingdom {\textendash} Department of Physics, University of Oxford.

\bibliographystyle{aasjournal}
\bibliography{clean_bib}

\end{document}

%% file: authors.tex



\affiliation{III. Physikalisches Institut, RWTH Aachen University, D-52056 Aachen, Germany}
\affiliation{Department of Physics, University of Adelaide, Adelaide, 5005, Australia}
\affiliation{Dept. of Physics and Astronomy, University of Alaska Anchorage, 3211 Providence Dr., Anchorage, AK 99508, USA}
\affiliation{Dept. of Physics, University of Texas at Arlington, 502 Yates St., Science Hall Rm 108, Box 19059, Arlington, TX 76019, USA}
\affiliation{CTSPS, Clark-Atlanta University, Atlanta, GA 30314, USA}
\affiliation{School of Physics and Center for Relativistic Astrophysics, Georgia Institute of Technology, Atlanta, GA 30332, USA}
\affiliation{Dept. of Physics, Southern University, Baton Rouge, LA 70813, USA}
\affiliation{Dept. of Physics, University of California, Berkeley, CA 94720, USA}
\affiliation{Lawrence Berkeley National Laboratory, Berkeley, CA 94720, USA}
\affiliation{Institut f{\"u}r Physik, Humboldt-Universit{\"a}t zu Berlin, D-12489 Berlin, Germany}
\affiliation{Fakult{\"a}t f{\"u}r Physik {\&} Astronomie, Ruhr-Universit{\"a}t Bochum, D-44780 Bochum, Germany}
\affiliation{Universit{\'e} Libre de Bruxelles, Science Faculty CP230, B-1050 Brussels, Belgium}
\affiliation{Vrije Universiteit Brussel (VUB), Dienst ELEM, B-1050 Brussels, Belgium}
\affiliation{Dept. of Physics, Massachusetts Institute of Technology, Cambridge, MA 02139, USA}
\affiliation{Dept. of Physics and Institute for Global Prominent Research, Chiba University, Chiba 263-8522, Japan}
\affiliation{Dept. of Physics and Astronomy, University of Canterbury, Private Bag 4800, Christchurch, New Zealand}
\affiliation{Dept. of Physics, University of Maryland, College Park, MD 20742, USA}
\affiliation{Dept. of Astronomy, Ohio State University, Columbus, OH 43210, USA}
\affiliation{Dept. of Physics and Center for Cosmology and Astro-Particle Physics, Ohio State University, Columbus, OH 43210, USA}
\affiliation{Niels Bohr Institute, University of Copenhagen, DK-2100 Copenhagen, Denmark}
\affiliation{Dept. of Physics, TU Dortmund University, D-44221 Dortmund, Germany}
\affiliation{Dept. of Physics and Astronomy, Michigan State University, East Lansing, MI 48824, USA}
\affiliation{Dept. of Physics, University of Alberta, Edmonton, Alberta, Canada T6G 2E1}
\affiliation{Erlangen Centre for Astroparticle Physics, Friedrich-Alexander-Universit{\"a}t Erlangen-N{\"u}rnberg, D-91058 Erlangen, Germany}
\affiliation{Physik-department, Technische Universit{\"a}t M{\"u}nchen, D-85748 Garching, Germany}
\affiliation{D{\'e}partement de physique nucl{\'e}aire et corpusculaire, Universit{\'e} de Gen{\`e}ve, CH-1211 Gen{\`e}ve, Switzerland}
\affiliation{Dept. of Physics and Astronomy, University of Gent, B-9000 Gent, Belgium}
\affiliation{Dept. of Physics and Astronomy, University of California, Irvine, CA 92697, USA}
\affiliation{Karlsruhe Institute of Technology, Institut f{\"u}r Kernphysik, D-76021 Karlsruhe, Germany}
\affiliation{Dept. of Physics and Astronomy, University of Kansas, Lawrence, KS 66045, USA}
\affiliation{SNOLAB, 1039 Regional Road 24, Creighton Mine 9, Lively, ON, Canada P3Y 1N2}
\affiliation{Department of Physics and Astronomy, UCLA, Los Angeles, CA 90095, USA}
\affiliation{Department of Physics, Mercer University, Macon, GA 31207-0001}
\affiliation{Dept. of Astronomy, University of Wisconsin, Madison, WI 53706, USA}
\affiliation{Dept. of Physics and Wisconsin IceCube Particle Astrophysics Center, University of Wisconsin, Madison, WI 53706, USA}
\affiliation{Institute of Physics, University of Mainz, Staudinger Weg 7, D-55099 Mainz, Germany}
\affiliation{Department of Physics, Marquette University, Milwaukee, WI, 53201, USA}
\affiliation{Institut f{\"u}r Kernphysik, Westf{\"a}lische Wilhelms-Universit{\"a}t M{\"u}nster, D-48149 M{\"u}nster, Germany}
\affiliation{Bartol Research Institute and Dept. of Physics and Astronomy, University of Delaware, Newark, DE 19716, USA}
\affiliation{Dept. of Physics, Yale University, New Haven, CT 06520, USA}
\affiliation{Dept. of Physics, University of Oxford, Parks Road, Oxford OX1 3PU, UK}
\affiliation{Dept. of Physics, Drexel University, 3141 Chestnut Street, Philadelphia, PA 19104, USA}
\affiliation{Physics Department, South Dakota School of Mines and Technology, Rapid City, SD 57701, USA}
\affiliation{Dept. of Physics, University of Wisconsin, River Falls, WI 54022, USA}
\affiliation{Dept. of Physics and Astronomy, University of Rochester, Rochester, NY 14627, USA}
\affiliation{Oskar Klein Centre and Dept. of Physics, Stockholm University, SE-10691 Stockholm, Sweden}
\affiliation{Dept. of Physics and Astronomy, Stony Brook University, Stony Brook, NY 11794-3800, USA}
\affiliation{Dept. of Physics, Sungkyunkwan University, Suwon 16419, Korea}
\affiliation{Dept. of Physics and Astronomy, University of Alabama, Tuscaloosa, AL 35487, USA}
\affiliation{Dept. of Astronomy and Astrophysics, Pennsylvania State University, University Park, PA 16802, USA}
\affiliation{Dept. of Physics, Pennsylvania State University, University Park, PA 16802, USA}
\affiliation{Dept. of Physics and Astronomy, Uppsala University, Box 516, S-75120 Uppsala, Sweden}
\affiliation{Dept. of Physics, University of Wuppertal, D-42119 Wuppertal, Germany}
\affiliation{DESY, D-15738 Zeuthen, Germany}

\author{M. G. Aartsen}
\affiliation{Dept. of Physics and Astronomy, University of Canterbury, Private Bag 4800, Christchurch, New Zealand}
\author{M. Ackermann}
\affiliation{DESY, D-15738 Zeuthen, Germany}
\author{J. Adams}
\affiliation{Dept. of Physics and Astronomy, University of Canterbury, Private Bag 4800, Christchurch, New Zealand}
\author{J. A. Aguilar}
\affiliation{Universit{\'e} Libre de Bruxelles, Science Faculty CP230, B-1050 Brussels, Belgium}
\author{M. Ahlers}
\affiliation{Niels Bohr Institute, University of Copenhagen, DK-2100 Copenhagen, Denmark}
\author{M. Ahrens}
\affiliation{Oskar Klein Centre and Dept. of Physics, Stockholm University, SE-10691 Stockholm, Sweden}
\author{C. Alispach}
\affiliation{D{\'e}partement de physique nucl{\'e}aire et corpusculaire, Universit{\'e} de Gen{\`e}ve, CH-1211 Gen{\`e}ve, Switzerland}
\author{K. Andeen}
\affiliation{Department of Physics, Marquette University, Milwaukee, WI, 53201, USA}
\author{T. Anderson}
\affiliation{Dept. of Physics, Pennsylvania State University, University Park, PA 16802, USA}
\author{I. Ansseau}
\affiliation{Universit{\'e} Libre de Bruxelles, Science Faculty CP230, B-1050 Brussels, Belgium}
% ...
\date{\today}

\collaboration{450}{IceCube Collaboration}
\noaffiliation
\email{analysis@icecube.wisc.edu}

%% file: event_numbers_table.tex
\begin{deluxetable*}{cccccc}




\tablecaption{Data Sample Information}
\savetablenum{1}

\label{table:event_numbers_table}

\tablehead{\colhead{Data Year} & \colhead{Livetime [Days]} & \colhead{$\lambda_s$ [m]} & \colhead{N$_{\mathrm{events}}$ (total)} & \colhead{N$_{\mathrm{events}}$ (PS)} & \colhead{N$_{\mathrm{events}}$ (GP)}}


\startdata
2011 & 308.7 & 2.10 & 27551210 & 97034 & 68286 \\
2012 & 295.9 & 2.25 & 35662684 & 85079 & 64823 \\
2013 & 321.4 & 2.25 & 35215316 & 107009 & 79787 \\
2014 & 325.7 & 2.30 & 33174803 & 96682 & 73473 \\
2015 & 325.2 & 2.30 & 30244777 & 85657 & 61907 \\
\enddata


\tablecomments{For each data year, the livetime of the data runs used in the final analysis, the snow absorption length used in the charge correction, as well as the number of events before classification, classified as gamma-ray candidates by the point source (PS) event selection, and classified as gamma-ray candidates by the galactic plane (GP) event selection.}


\end{deluxetable*}

%% file: hess_table.tex


\begin{deluxetable*}{lccc|cc|cc}


\tablecaption{H.E.S.S. Source Target Catalog}
\savetablenum{2}
\label{table:hess_p_values}

\tablehead{\colhead{Source} & \colhead{Type} & \colhead{Declination} & \colhead{Spectral Index} & \colhead{p-value} & \colhead{$\Phi_{90\%}$ (2 PeV)} & \colhead{Distance} & \colhead{E$_{\textrm{cut}}$ U.L.}\\ \colhead{} &  \colhead{} & \colhead{[$^{\circ}$]} & \colhead{} & \colhead{} & \colhead{[\si{cm^{-2}s^{-1}TeV^{-1}}]} & \colhead{[kpc]} & \colhead{[PeV]}}

\startdata
HESS J1356-645$^1$   & PWN & -64.50 & 2.20 & $>$0.50  & \num{7.66e-21} & 2.4$^1$  & 0.9   \\
HESS J1507-622$^2$   & PWN & -62.35 & 2.24 & 0.29     & \num{1.42e-20} & 8.5* & 16.6\\
HESS J1119-614$^{3,\textrm{A}}$   & Unidentified & -61.40 & N/A  & 0.30     & N/A            & \phd5.0$^{13}$  & N/A         \\
HESS J1418-609$^4$  & SNR & -60.98 & 2.22 & 0.31     & \num{1.34e-20} & 5.0$^4$  & 3.5    \\
HESS J1458-608$^5$  & Unidentified & -60.87 & 2.80 & 0.22     & \num{2.56e-20} & 8.5* & ---     \\
HESS J1427-608$^6$  & PWN & -60.85 & 2.20 & 0.07     & \num{1.96e-20} & 8.5* & 25.9 \\
HESS J1420-607$^4$  & PWN & -60.76 & 2.17 & 0.49     & \num{8.84e-21} & \phd5.6$^{14}$  & 1.4    \\
HESS J1457-593$^{7,\textrm{A}}$   & SNR & -59.47 & N/A  & $>$0.50  & N/A            & 8.5* & N/A     \\
HESS J1514-591$^8$   & PWN & -59.16 & 2.27 & 0.52     & \num{1.02e-20} & \phd5.2$^{15}$  & 1.5    \\
HESS J1018-589 B$^9$ & PWN & -58.98 & 2.90 & 0.08     & \num{4.12e-20} & 8.5* & ---     \\
HESS J1018-589 A$^9$ & Unidentified & -58.93 & 2.20 & 0.07     & \num{2.17e-20} & 8.5* & ---     \\
HESS J1503-582$^{10}$ & Unidentified & -58.23 & 2.40 & 0.17     & \num{2.77e-20} & 8.5* & ---     \\
HESS J1026-582$^{11}$ & PWN & -58.20 & 1.94 & 0.09     & \num{1.83e-20} & \phd2.3$^{16}$  & 1.4    \\
HESS J1023-575$^{11}$ & Unidentified & -57.79 & 2.58 & 0.08     & \num{4.60e-20} & \phd8.0$^{17}$  & ---         \\
HESS J1554-550$^{12}$ & PWN & -55.08 & 2.10 & $>$0.50  & \num{2.29e-20} & \phd9.0$^{18}$  & ---         \\
\enddata

\tablecomments{First, the source type, declination, and best-fit spectral index by H.E.S.S. for each candidate source.  Next, the pre-trial p-value observed by this analysis and 90\% confidence level upper limits on the flux at 2 PeV at Earth assuming the H.E.S.S. best-fit spectral index and no cut-off in energy.  Finally, the best known distance to each source and a 90\% confidence level upper limit on the energy cut-off of the source assuming an extrapolation of the H.E.S.S. best-fit power law spectrum with an energy cut off and absorption from~\citet{Vernetto:2017}.  
Sources where this analysis cannot limit the energy cut off are denoted by ---.}
\vspace{2mm}
\,$\enspace^{\textrm{A}}$Sources with no reported spectral index, in which case we do not report source specific limits.\\
\vspace{2mm}\\
*Sources without an x-ray or radio observation are given a value of 8.5 kpc, approximately the distance to the Galactic center.

\tablerefs{$^1$\citet{Abramowski:2011}, $^2$\citet{Acero:2011}, $^3$\citet{Djannati:2012}, $^4$\citet{Aharonian:2006b}, $^5$\citet{Reyes:2012}, $^6$\citet{Aharonian:2008}, $^7$\citet{Hofverberg:2010}, $^8$\citet{Aharonian:2005b}, $^9$\citet{Abramowski:2012}, $^{10}$\citet{Renaud:2008}, $^{11}$\citet{Abramowski:2011b}, $^{12}$\citet{Acero:2011b}, $^{13}$\citet{Crawford:2001}, $^{14}$\citet{Kishishita:2012}, $^{15}$\citet{Gaensler:1999}, $^{16}$\citet{Keith:2008}, $^{17}$\citet{Rauw:2011}, $^{18}$\citet{Sun:1999}}



\end{deluxetable*}